\documentclass[aps,twocolumn,pra,showpacs,amsmath,nofootinbib]{revtex4}
\usepackage{graphicx}
\usepackage{dcolumn}
\usepackage{bm}
\usepackage{epstopdf}
\newcommand{\beq}{\begin{equation}}
\newcommand{\eeq}{\end{equation}}
\newcommand{\bea}{\begin{eqnarray}}
\newcommand{\eea}{\end{eqnarray}}

\begin{document}

\author{A. Avdeenkov}
\affiliation{ National Institute for Theoretical Physics,
Stellenbosch Institute of Advanced Study, South Africa;\\
 Institute of Nuclear Physics, Moscow State University,
Vorob'evy Gory, Moscow, Russia}
\author{S. Goriely}
\affiliation{Institut d'Astronomie et d'Astrophysique,
ULB,  CP 226, B-1050 Brussels, Belgium}
\author{S. Kamerdzhiev}
\affiliation{Institute of Physics and Power Engineering, 249033 Obninsk, Russia;\\
 Institut fuer Kernphysik, Forshungszentrum Juelich, D-52425 Juelich, Germany}
\author{S. Krewald}
\affiliation{Institut fuer Kernphysik, Forshungszentrum Juelich, D-52425 Juelich, Germany}

\title{Self-consistent calculations of the strength function and radiative neutron capture cross section
for stable and unstable tin isotopes}

\pacs{24.10.-i, 24.60.Dr, 24.30.Cz, 21.60.Jz }

\begin{abstract}
The E1 strength function  for 15 stable and unstable Sn even-even isotopes from $A=100$ till $A=176$
are calculated using the  self-consistent  microscopic theory which, in addition to the standard  (Q)RPA approach,  takes into account the single-particle continuum and the phonon coupling. Our analysis shows two distinct regions for which the integral characteristics of both the giant and pygmy resonances behave rather differently. For neutron-rich nuclei, starting from $^{132}$Sn, we obtain a giant E1 resonance which significantly deviates from the widely-used systematics extrapolated from experimental data in the valley of $\beta$-stability. We show that the inclusion of the phonon coupling is necessary for a proper description of  the low-energy pygmy resonances and the corresponding  transition densities for A$<$132 nuclei,  while in the $A>132$ region  the influence of the phonon coupling  is significantly smaller. The radiative neutron capture cross sections leading to the stable $^{124}$Sn and unstable $^{132}$Sn, $^{150}$Sn nuclei are calculated  with both the (Q)RPA and the beyond-(Q)RPA strength functions and shown to be sensitive  to both the predicted low-lying strength and the phonon coupling contribution.
The comparison with the  widely-used phenomenological Generalized Lorentzian approach shows considerable differences both  for  the strength function and the radiative neutron capture cross section. In particular, for the neutron-rich $^{150}$Sn, the reaction cross section is found to be increased by a factor larger than 20. We conclude that the present approach may provide a complete and coherent description of the $\gamma$-ray strength function for astrophysics applications. In particular, such calculations are highly recommended for  a reliable estimate of the electromagnetic properties of exotic nuclei.
\end{abstract}
\maketitle

\section{Introduction}
One of the paramount and challenging  goals of modern nuclear physics is to elaborate theoretical approaches with not only descriptive but also predictive abilities. This is of particular relevance for a proper description of nuclei far from the valley of stability since in this case only limited or no information is available.
Self-consistency between the mean field and the effective interaction is also known to be of prime importance for a correct exploration of the excitation properties of unstable nuclei.
Another fundamental ingredient of the model, known to be important even for stable nuclei, concerns the inclusion of more complex configurations than those traditionally included in the Random Phase Approximation (RPA) or Quasi-particle RPA (QRPA).
Here the most realistic approaches include  complex configurations  with phonons, i.e. the coupling of single-particle degrees of freedom with the phonon degrees (the so-called phonon coupling or PC).
These approaches are referred in the litterature as the Quasi-particle-Phonon Model \cite{VG}, the (Q)RPA+Phonon Coupling Model ((Q)RPA+PC) \cite{colo94} and the Extended Theory of Finite Fermi Systems (ETFFS) \cite{ETFFS2004}. The latter is based on the Green function method and includes the single-particle continuum which is necessary for nuclei with a nucleon separation energy close to zero. It has been recently generalized to include pairing using the quasi-particle time blocking approximation (QTBA) \cite{tse07}.

These approaches have been supplemented by considering a self-consistent mean field, see for example \cite{tsoneva}, or the self-consistency between the mean field and the effective interaction~\cite{sarchi,litvaring1,avd07}. The latter made it possible to perform the calculation with one unique set of interaction parameters, for example the Skyrme force~\cite{sarchi,avd07}, instead of  two sets of  parameters used in non-self-consistent approaches (one for the effective interaction and another for the mean field).
These improvements - taking into account the single-particle continuum and self-consistency- are of great interest, first of all, for astrophysics applications, but also for nuclear data evaluation. The added-value of such approaches lies essentially in their larger predictive power which provides an increased confidence in the calculation of  structure properties for exotic nuclei, especially those with a large neutron excess and /or a small nucleon separation energy. However, one should note that all the approaches developed so far are in fact not fully self-consistent because, as it was noted earlier in \cite{kaev1,kaev2}, they use the self- consistency only at the  (Q)RPA level and do not include  more complex configurations into the self-consistency conditions.
This is one of the main reasons to use  some additional procedures  to exclude  ghost states, in particular,
the spurious isoscalar 1$^-$ state. This is achieved through a specific fit of the force parameters \cite{tsoneva,sarchi,avd07}, the use of  the so-called subtraction procedure in the QTBA model \cite{tse07} or of  the so-called ``forced consistency''  method \cite{kaevliotta1998,littsel2007}.
Nevertheless, accounting for  the PC and self-consistency,  beyond any doubt, increases the quality of the microscopic nuclear theory and is absolutely necessary to describe simultaneously the structure of ground and excited states for unstable nuclei.

The role played by the PC in the description of the giant resonances \textit{in stable nuclei} is well-known. In particular,  the  PC explains approximately $50\%$ of the observed width, its  gross structure  and sometimes  some fine structure (e.g. for the E2 isoscalar resonance in $^{208}$Pb \cite{ETFFS2004}). However, the direct influence of the PC on  the giant resonance of unstable nuclei has been much less studied though it is expected to be as important as it is for stable nuclei. So far, giant resonance characteristics for unstable nuclei have been studied systematically  within the (Q)RPA only (for example, see \cite{rev,gor02,gor04,sagawa})  and, quite recently,  within a general approach based on sum rules \cite{colo2009}. Note that for the reasons given below,  we will discuss in the present paper only the case of electric dipole resonance.

The impact of the PC on the so-called pygmy dipole resonance (PDR), which lies in the low-energy tail of the E1 giant dipole resonance (GDR) and exhausts about $1-2\%$ of the energy-weighted sum rule (EWSR)~\cite{kneisl,volz,gor02,ripl2}, is of particular interest. First, there is no consensus at present  in our understanding of some important questions related to the PDR (see \cite{rev} and a recent mini-review \cite{avekaev2009}).
Second, this resonance is known to have a significant impact on the radiative neutron capture rate of astrophysical interest~\cite{gor98,gor02,gor04}. The importance of the PDR is confirmed  by the simple fact that it has been taken into account in all modern nuclear data libraries (though at a phenomenological level) in addition to the usual GDR  \cite{ripl2,capote09}.
The question arises for exotic nuclei where the phenomenological approach may fail to provide a reliable prediction because of the specific features of the PDR in such nuclei and the scarce experimental data on which systematics is based. For this reason, a similar approach should be followed as for the giant resonance problem for unstable nuclei,  i.e.,  as discussed above, to use a reliable self-consistent theory which accounts for PC and the single-particle continuum in addition to the standard (Q)RPA.

For all these reasons, the PDR problem has recently become a subject of intensive experimental (see Refs.~\cite{kneisl,volz,wieland} and references therein) and theoretical (see Refs. \cite{suzuki1990,sarchi,tsoneva,tert07,litvaring1,avdkoeln2008,avekaev2009} and references therein) studies.
Even if the total E1 strength of the PDR is small, if located well below the neutron separation energy,  it can significantly increase the radiative neutron capture cross section, especially, for neutron-rich nuclei \cite{gor98,gor02,gor04}. Different measurements suggest that some enhancement of the E1-strength could be located at low energies even on stable nuclei, a feature that cannot be described within the (Q)RPA calculations.
In particular, the large-scale QRPA calculations of \cite{gor02} predict  PDRs which are  on average 1 to 2~MeV
higher in energy than the observed values. Many recent calculations of the PDR \cite{sarchi,litvaring1,avdkoeln2008,avekaev2009,colo2001,kaevlit2004} as well as the older ones \cite{ponomarev1, ponomarev2} performed within the non-self-consistent Quasi-particle-Phonon Model  confirm the need to take into account more complex configurations than those included in the (Q)RPA approach, most of all the 1p1h $\otimes$ phonon or 2 quasi-particles $\otimes$ phonon configurations. However, large uncertainties in the description of the PDR (in particular, its energy and strength) remain, especially for unstable nuclei, and only sound microscopic models can shed light on its existence, as well as its relative importance and impact on neutron capture. For example, the self-consistent calculations with PC \cite{avdkoeln2008,avddubna2009}  have shown that the complex configurations give a significant contribution to the radiative neutron capture cross section for the unstable $^{132}$Sn.

In practice, for a proper description of the PDR, at least two natural physical conditions need to be fulfilled: first, the energy of the 1$^-$ spurious state must be equal to zero; second, the theory must describe correctly the mean energy E$_0$ of the E1 giant resonance.  Only in this case one may expect  the theory to provide a reasonable quantitative prediction of the PDR integral features. In order to satisfy these two conditions different additional procedures have been used. The simplest way is to adjust the isovector and isoscalar effective  force parameters to obtain the correct values of E$_0$ and the spurious 1$^-$ level energy~\cite{tsoneva}. This is suitable for stable nuclei for which E$_0$ is experimentally available. For unstable nuclei,  if use is made of a Skyrme force and a self-consistent scheme without the subtraction procedure \cite{tse07}, it is necessary to modify some of the Skyrme
parameters  to obtain an agreement with experiment  \cite{avd07}. The first attempt to take the single-particle continuum into account exactly at the RPA+PC level using the Green function method was made in Ref.~\cite{lyutor08} for magic nuclei. It was shown that a renormalization of the SLy4 force was necessary to obtain an agreement with experiment. It is worth noting that  this conclusion is in accordance  with the studies of \cite{bender,dobach} who considered this idea from a different point of view.

In our previous works \cite{avd07,avdkoeln2008,avekaev2009} we realized a self-consistent version of the ETFFS(QTBA) using a discretized single-particle continuum with different kinds of Skyrme forces
including the SLy4 one, where the velocity force is considered in a local approximation (sometimes we call it DTBA). The latter had consequences for the renormalization of the interaction in order to locate the spurious state at zero energy. On the one hand, it is of great interest to obtain a general information about the E1 strength function and correspondingly about the radiative neutron capture cross section for many  neutron-rich nuclei using the well-known SLy4 forces. On the other hand, it is clear that the inclusion of the single-particle continuum along with the PC effects for non-magic nuclei is still a  difficult problem.  For these reasons, we use here our DTBA approach to calculate the E1 strength function for the  long Sn isotopic chain.

The aim of the present work is twofold. First, we calculate the PDR and GDR in the long chain of the stable and unstable tin isotopes using the variant of the microscopic self-consistent version of the ETFFS(QTBA) which, in addition to the (Q)RPA approach, takes into account the single-particle continuum (by means of a discretization procedure) and phonon coupling in nuclei with pairing. For this part we concentrate on the description of the integral characteristics in order to gain an insight view into different trends for stable and unstable nuclei and to compare our results with the widely-used empirical formula. Second, in order to investigate the impact of the PC on the radiative neutron capture cross section in stable and unstable nuclei we calculate them both with and without PC, within the same scheme of calculation based on the  SLy4 Skyrme force or a slightly modified version of it. Here our main attention is paid to the PDR and its impact on the radiative neutron capture.

\section{Self-consistent  calculation of the PDR and GDR}

\subsection{Method}
To date there are tens of different Skyrme parametrizations serving slightly different  aims and fitting some bulk properties of the ground state. Here we use the SLy4 parametrization of  the Skyrme force  \cite{SLy4} which proves to be rather successful in describing bulk properties of the ground state and some excited states within the (Q)RPA~\cite{terasaki1}.

The ground states are calculated within the  HFB approach using the spherical code HFBRAD~\cite{bennaceur}. The residual interaction for the (Q)RPA and  QTBA calculations  is derived as the second derivative of the Skyrme functional~\cite{terasaki1}. In our calculation, several simplifications are performed. Namely, since up to now the QTBA approach is designed to use the BCS-based quasi-particle basis,  we use the HFB approach to extract the quasi-particle
characteristics and corresponding wave functions, i.e. the occupation numbers are treated as in the BCS approximation. The spin-orbit residual interaction is dropped. The velocity-dependent terms of the Skyrme force are approximated by their Landau-Migdal limit~\cite{speth,krewald1977} though some physically sounder modifications are included. There are two kinds of velocity-dependent terms: the first one is $\propto \bf{k^2} \delta (r-r')$ and the second one to $\bf{k^\dag} \delta (r-r') \bf{k}$ (P-wave interaction in the momentum space). The averaged value  over the density of the first term gives $k_F^2/2 \delta(r-r')$ while that of the second one is zero. Such an approximation violates the self-consistency and one has to correct the parameters of the residual interaction to put the spurious center-of-mass state to zero. We only change here the term which is proportional to $t_1 \bf{k_F^2} \delta (r-r')$ by a given factor as we take this term approximately. This factor is usually around $1.0-1.25$ for the Sn chain.

In general, the ETFFS(QTBA) accounts for the single-particle continuum completely at the RPA level for magic nuclei  and includes the new effect of ground state correlations caused by the PC~\cite{ETFFS2004}. However, because of the technical difficulties connected with the pairing specificity,  these effects are not considered in the present calculations. The quasi-particle energy cutoff of 100~MeV is used.
We checked that within this approach the EWSR is fully exhausted (for the case without the velocity-dependent terms) and that the use of a larger basis did not bring any noticeable differences. The QTBA calculations  are performed with the same basis. We use 14-16 low-lying phonons of $L=2-6$ multipolarity and normal parity. They are obtained within the (Q)RPA  with the calculated effective interaction using the same quasi-particle-energy cutoff. Such a consistent method to calculate phonons is the reason for us to use a larger number of phonons than in the phenomenological ETFFS \cite{ETFFS2004}. In  Fig.~\ref{fig:st2} we test our numerical approximation of  the single-particle continuum discretization for the $^{132}$Sn and $^{176}$Sn magic nuclei, by comparing our RPA results with  the exact account for the continuum by the Green function method at the RPA level~\cite{shlomo,kiae}, as described in Ref.~\cite{lyutor08}. It turns out that both calculations are almost identical which confirms that the discretization procedure adopted here is quite satisfactory.

Usually the GDR strength function  is
obtained from the experimental photoabsorption cross section which is fitted by a simple Lorentzian functional. The E1 photoabsorption cross section is related to the strength function $S(\omega)$  as follows
$
\sigma_{E1}(\omega) = 4.022~\,\omega~S_{E1}(\omega),
$
where the photon energy $\omega$ is taken in MeV, $S$ in fm$^{2}$~MeV$^{-1}$ and
$\sigma$ in mb.
The Lorentzian fit can be used to estimate the integral characteristics of the giant resonance~\cite{kst93,tselyaev00}. The mean energy E$_{0}$, the resonance width ${\Gamma}$ and the maximum value of the cross section $\sigma_{0}$ are extracted from the calculated photoabsorption cross sections under the condition that the three lowest energy-weighted moments of the Lorentzian and of the theoretical curve should coincide  in the considered energy interval. However, in order to
obtain a more complete and universal information about the integral characteristics it is often better to use energy moments (see below). This seems to be more appropriate  for very neutron-rich nuclei where one may expect significant deviations from a Lorentzian-like shape for the cross section. The same (0-30)~MeV  summation interval is used for all considered nuclei.

\begin{figure}
{\includegraphics[width=1.1\linewidth,height=1.0\linewidth,angle=-0]{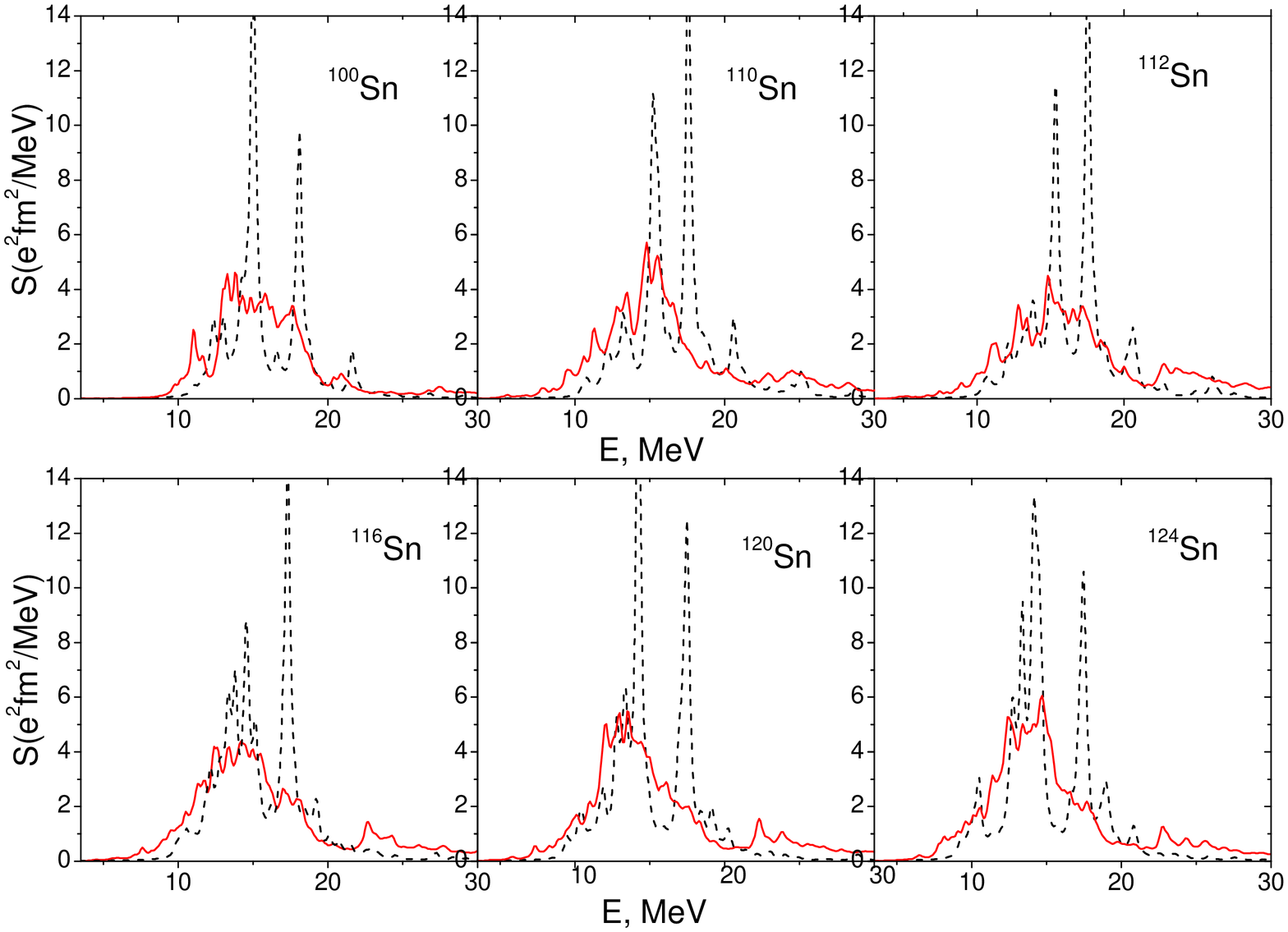}}
 \caption{(Color online) $1^{-}$ strength functions within QRPA (dashed curves) and QTBA (solid curves)  for $^{100}$Sn,$^{110}$Sn,$^{112}$Sn,$^{116}$Sn,
 $^{120}$Sn,$^{124}$Sn isotopes. }
 \label{fig:st1}
\end{figure}

\begin{figure}
{\includegraphics[width=1.1\linewidth,height=1.0\linewidth,angle=-0]{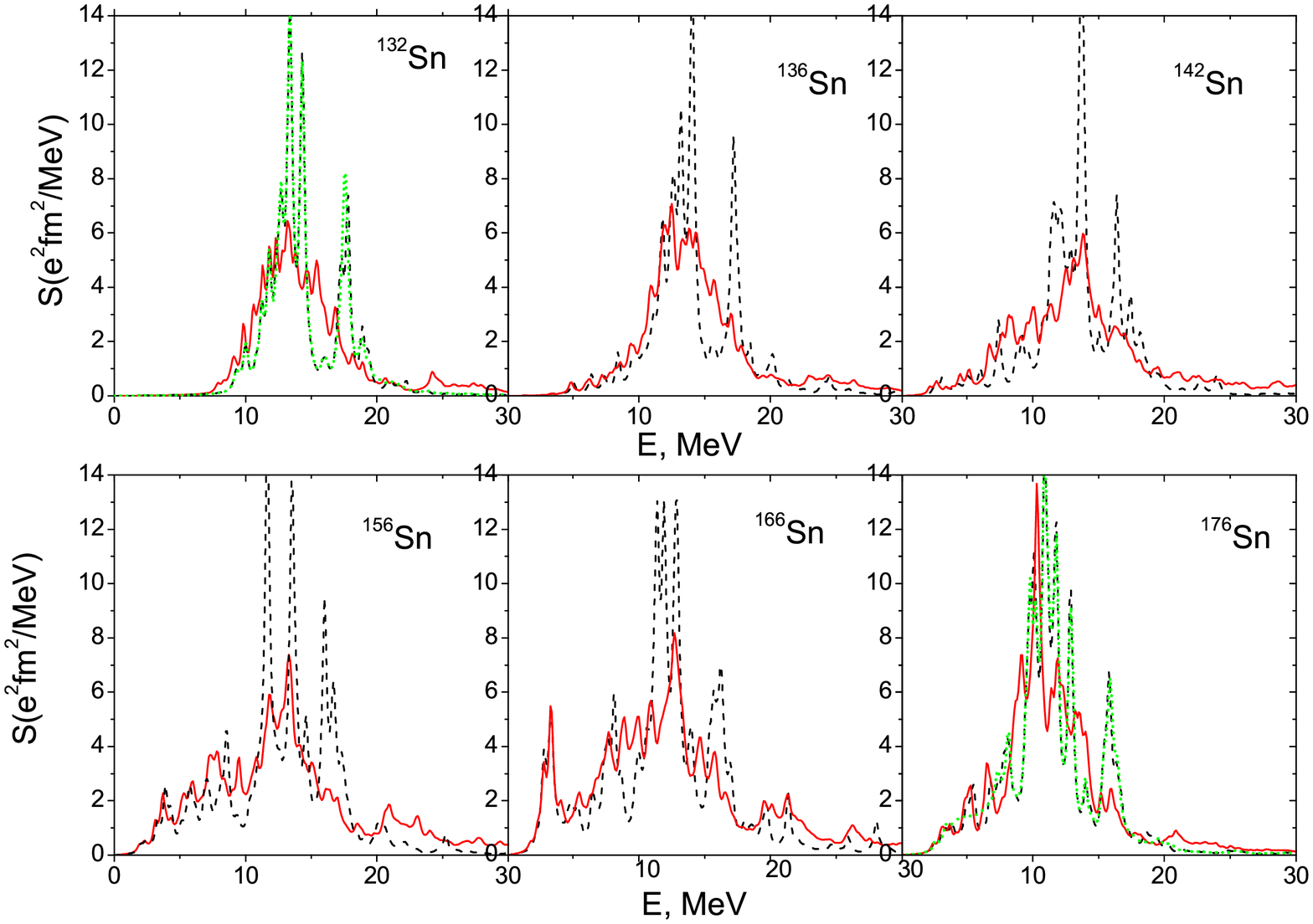}}
 \caption{(Color online) $1^{-}$ strength functions within QRPA (dashed curves) and QTBA (solid curves)  in $^{132}$Sn,$^{136}$Sn,$^{142}$Sn,$^{156}$Sn,$^{166}$Sn,
 $^{176}$Sn. For $^{132}$Sn and  $^{176}$Sn isotopes, the solid green curves show calculations within the continuum RPA.}
\label{fig:st2}
\end{figure}

The other feature of our calculation scheme is  the so-called subtraction procedure
\cite{tse07,litvaring1} which should
avoid a PC double counting  from the effective interaction. This procedure  is a direct continuation   of the phenomenological refinement philosophy used in the first formulations of the ETFFS \cite{yadfiz1983,ETFFS2004}.
Because the self-consistent relativistic (Q)RPA calculations are well fitted to experimental data  and because the subtraction procedure, in principle, provides the correct E$_0$ value, which is equal to the (Q)RPA $E_0$ value, the authors \cite{litvaring1} obtained a fulfillment of the above-mentioned conditions if  the energy interval is properly chosen, for example (10-22.5) MeV for Z=50 nuclei or (10-25) MeV for N=50 nuclei. However, altogether with a reasonable description of the GDR's width, the PC gives an additional strength in the low-energy region (see  Figs.\ref{fig:st1},\ref{fig:st2}). So, this low-energy contribution of the strength is usually neglected in such an  analysis of the integral characteristics based on the Lorentzian fit.

\begin{figure}
{\includegraphics[width=.95\linewidth,height=0.70\linewidth,angle=-0]{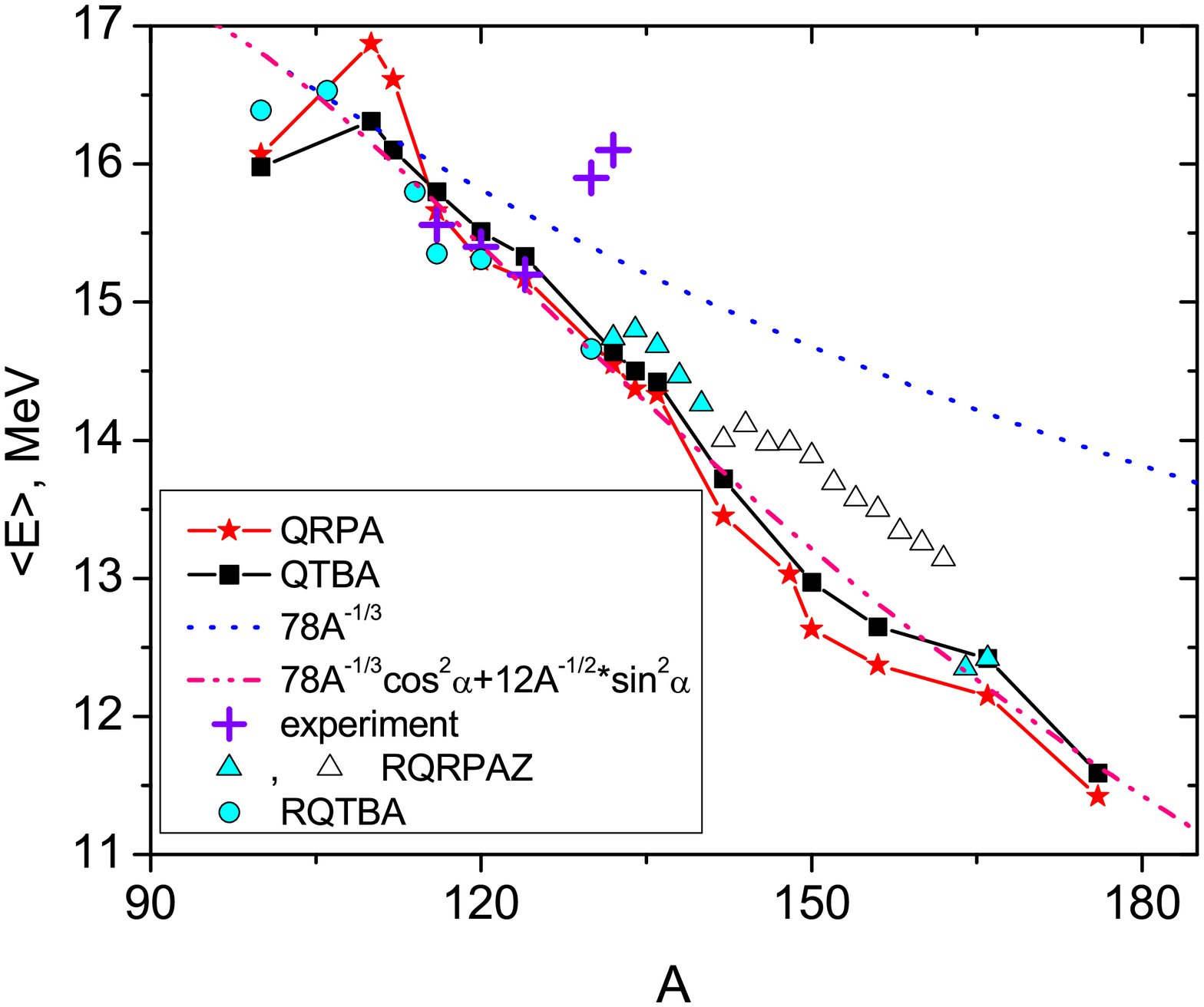}}
\hfill {\includegraphics[width=0.95\linewidth,height=0.70\linewidth,
angle=-0]{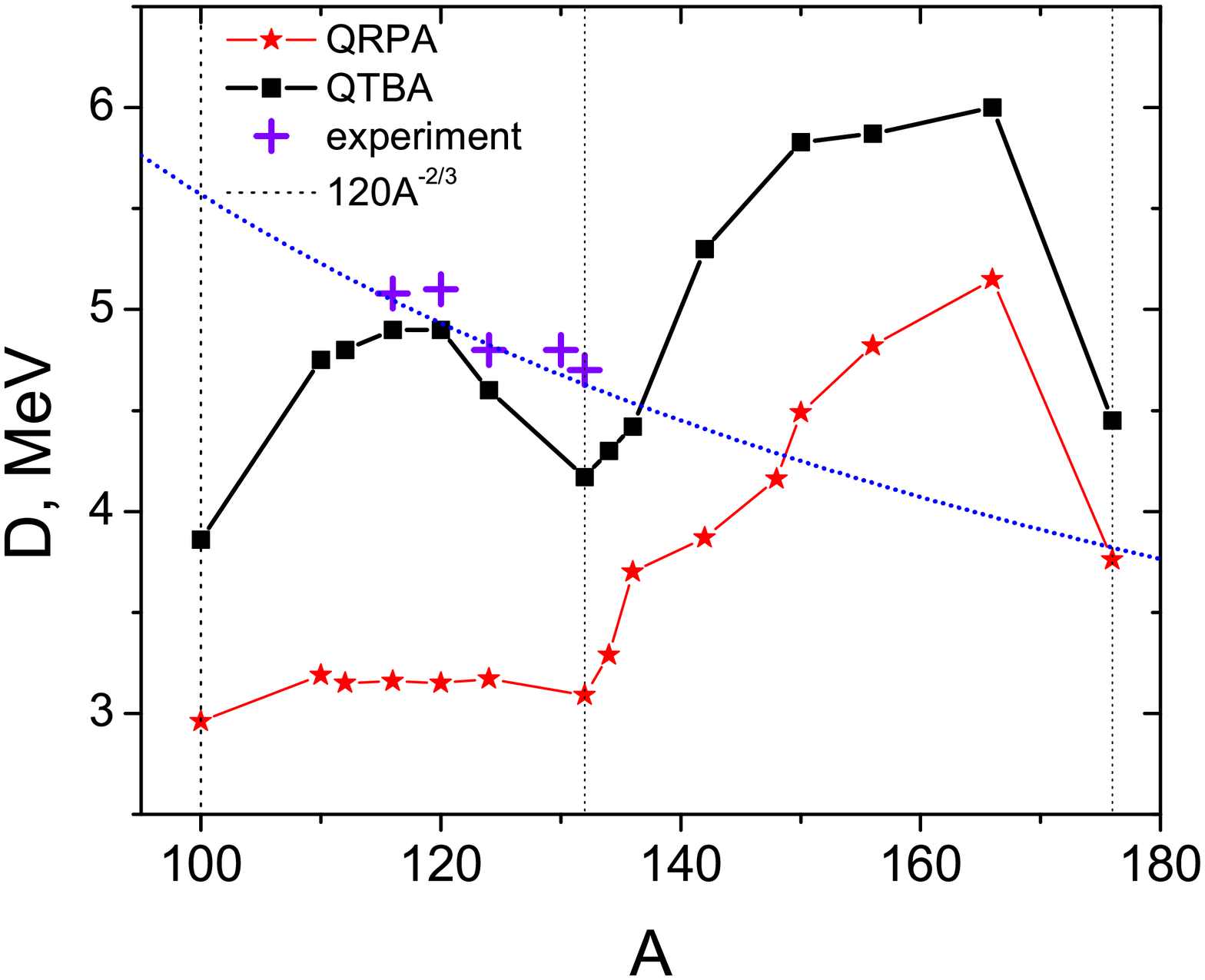}}
 \caption{(Color online) Integral characteristics of the $1^{-}$ state for some Sn isotopes:
{\it upper panel}- the GDR mean energy  versus the atomic mass number $A$ in the (0-30) MeV interval;
{\it lower panel}- the GDR dispersion  versus $A$ for the same interval.
 RQTBA and  RQRPAZ results are taken from Refs.~\cite{ring2009,litvaring1}, respectively.}
 \label{fig:in1}
\end{figure}

\subsection{Results}
Here we discuss our results for the long chain of tin isotopes from $^{100}$Sn up to $^{176}$Sn. Figs.\ref{fig:st1} and \ref{fig:st2} show the strength functions for twelve Sn isotopes. In the following subsections, the  dipole excitations are studied and the GDR and PDR analyzed along the Sn isotopic chain. The general idea behind such an investigation is to understand if there are  some common trends for dipole excitations in stable and unstable isotopes on both sides of the $\beta$-stability valley and if they can be described within one unique scheme and with one unique force like SLy4.  One can argue that such a force may not be appropriate for unstable species but this force is among the most suited tool and can provides us with valuable theoretical findings, for example, as the non-Lorentzian shape of the GDR for very neutron-rich nuclei such as $^{156}$Sn, $^{166}$Sn (see Fig.\ref{fig:st2}), as discussed below.

\subsubsection{Dipole excitations and Giant Dipole Resonances}
Fig.\ref{fig:in1} shows the calculated integral characteristics of the dipole excitations for fifteen stable and unstable Sn isotopes. The smearing parameter is $\Delta =200$~keV for all  calculations. The E1 response integral
characteristics for the mean energies and dispersions are  calculated   using the standard definitions
\begin{equation}
<E> = E_{1,0} = \frac{m_{1}}{m_{0}}, \hspace{3mm}
D = \sqrt{\frac{m_{2}}{m_{0}} - (\frac{m_{1}}{m_{0}})^{2}},
\end{equation}
where the energy moments $m_{k}$ for  the energy interval $\Delta E = E_{max} - E_{min}$  are calculated as follows
\begin{equation}
m_{k} = \int_{E_{min}}^{E_{max}}dE \; E^{k} S(E).
\end{equation}

First we check  our approach on the stable Sn isotopes ($^{116}$Sn, $^{120}$Sn, $^{124}$Sn) and obtain, as shown in Fig.\ref{fig:in1}, a reasonable agreement with available experimental data~\cite{ripl2}.
The dipole excitation for these nuclei has a well visible Lorentzian-like form with parameters which may slightly vary depending on the adopted model~\cite{ripl2}. Though our results are not fitted by a Lorentzian form and only moments are compared, it can be seen that we have a reasonable agreement for both the mean energy and the width of the E1 resonance. The $^{132}$Sn and $^{130}$Sn are the only unstable tin isotopes which were probed to study the GDR and PDR~\cite{aumann05}. Though the SLy4 forces works quite well for stable nuclei there is no proof that it is good for unstable ones. In Ref.~\cite{avd07} along with SLy4 Skyrme forces we probed BSk5~\cite{BSk5} and SkM*~\cite{SkM*} for $^{132}$Sn. In all the three cases we obtained very similar mean energy values  and widths (i.e $14.3$~MeV and $2.9$~MeV, respectively, for the SLy4 force), while the experimental data are $16.1\pm0.7$~MeV and $4.7\pm2.1$~MeV~\cite{aumann05}. It has to be noted that other theoretical approaches (see Fig.~\ref{fig:in1})~\cite{litvaring1,ring2009} give very similar results which possibly means that further experimental investigations on the $^{132}$Sn GDR may be needed.

Our calculations show a noticeable difference both between (Q)RPA and ETFFS(QTBA) approaches and stable and unstable nuclei~(Figs.~\ref{fig:st1}-\ref{fig:in1}). The results for  integral characteristics (Fig.~\ref{fig:in1}) clearly show the necessity to take the PC into account for a proper determination of the GDR width. For the $A=100-132$ nuclei, the PC gives rise to an increase of the width by as much as 2~MeV as compared with the (Q)RPA predictions. This PC effect is also important in $A>132$ nuclei though to  a lesser extent. Fig.~\ref{fig:in1} shows two distinct regions for the integral characteristics. The first region corresponds to the stable isotopes, $116<A<124$, for which the integral characteristics follow the well-known phenomenological systematics (e.g. $E_0 \propto A^{-1/3}$ and $\Gamma \propto A^{-2/3}$). The second region includes unstable isotopes, $A>132$, for which these systematics fail. Similar conclusions for mean energies can be made out of the calculations obtained within the  RQRPAZ approach~\cite{ring2009}.

Here  we suggest another empirical systematics which describes quite well the mean energy in both regions defined above:
\begin{equation}
E_0=78A^{-1/3}cos^{2}\alpha+12A^{-1/2}sin^{2}\alpha,
\label{eq:eph}
\end{equation}
where $\alpha =(N-Z)/A$ is the neutron excess and the factor $12A^{-1/2}$ describes empirically  the pairing gap, see \cite{ringsch}. One can see that this is in  accordance  with the results in Fig. 6 (see below). The first term is responsible  for the collective GDR while the second one reflects the leftovers of the GDR, namely the non-collective particle-hole excitations. This formula reflects the fact that the mean energy depends on the superposition of these two kinds of excitations, while the degree of mixing is defined by the neutron excess  $\alpha$.  The second term becomes important for unstable nuclei and is correlated with the increased low-lying E1 strength.  As mentioned above, the E1 strength can hardly  be described by a Lorentzian function over the whole isotopic chain. One may try also to separate out the GDR and PDR in  the same way as done in ~Ref.~\cite{ring2009} using $^{132}$Sn as the benchmark in the definition of the border between these two resonances.
This procedure is somewhat artificial and anyway  for unstable nuclei the GDR centroid energy itself does not follow the systematics, nor does the mean energy taken in the whole $(0-30)$~MeV interval. Another interesting feature is that the
dispersion~(Fig.\ref{fig:in1}, lower panel) is minimal at neutron magic numbers (N=50, 82, 126) and maximal  for open shell nuclei. Likewise for the mean energy, the dispersion is quite close to the lorentzianian width for stable nuclei only, while the contribution of the low-lying tail to the dispersion for the second region is increasingly important with the increasing of the neutron excess.

\subsubsection{Pygmy Dipole Resonances}
\begin{figure}
\label{st1}
{\includegraphics[width=.88\linewidth,height=0.65\linewidth,angle=-0]{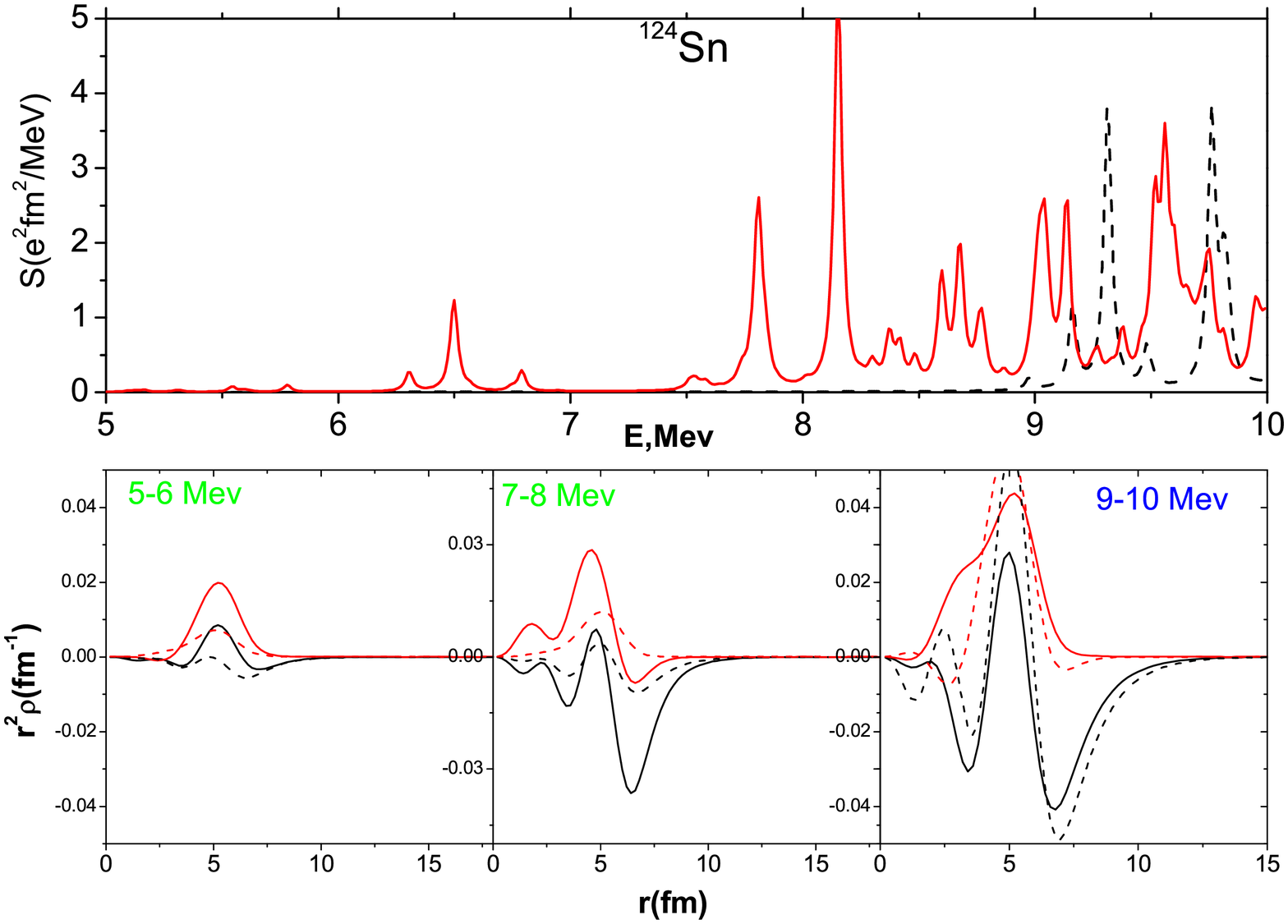}}
\hfill {\includegraphics[width=.88\linewidth,height=0.65\linewidth,
angle=-0]{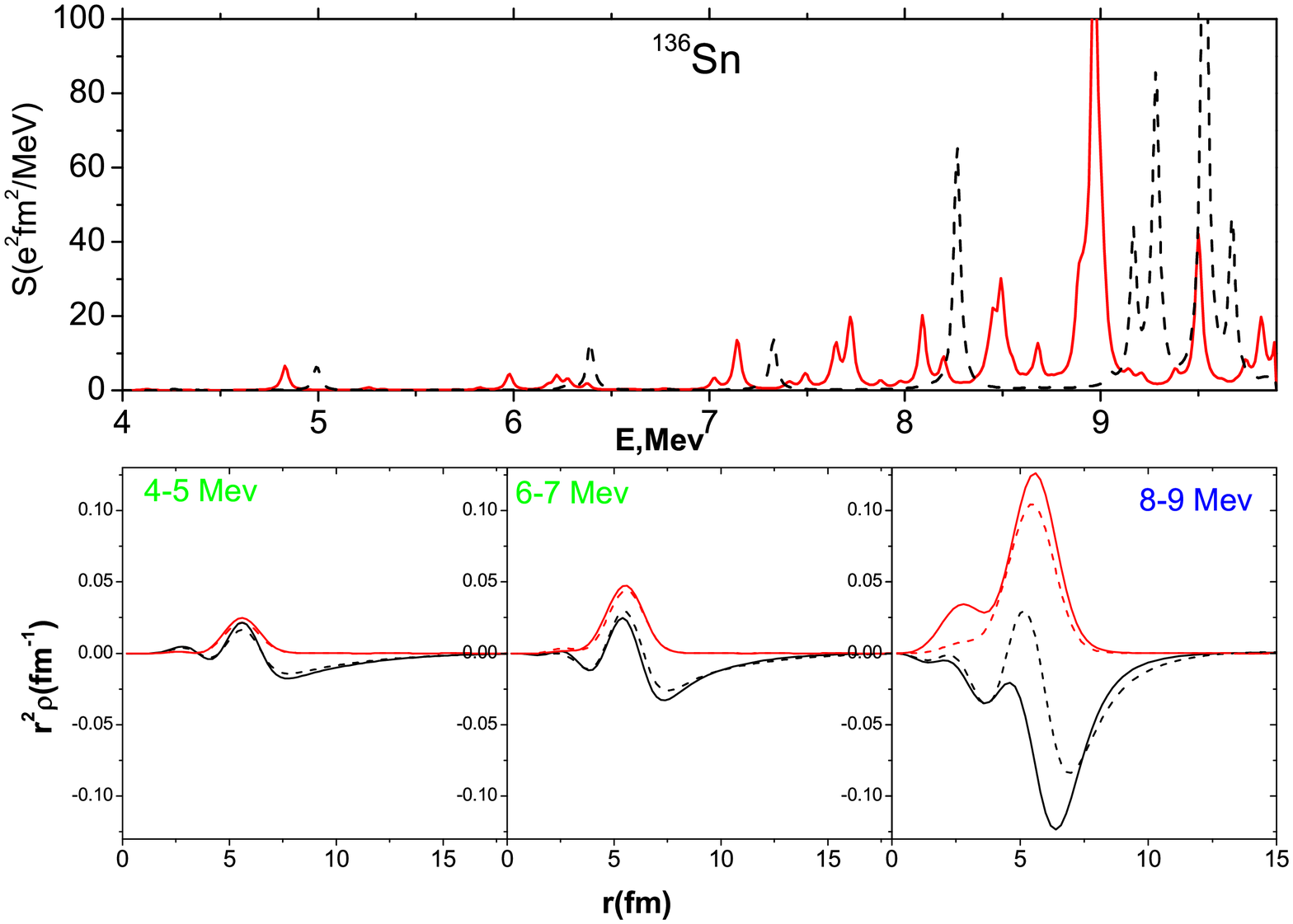}}
\hfill
{\includegraphics[width=0.88\linewidth,height=0.65\linewidth,
angle=-0]{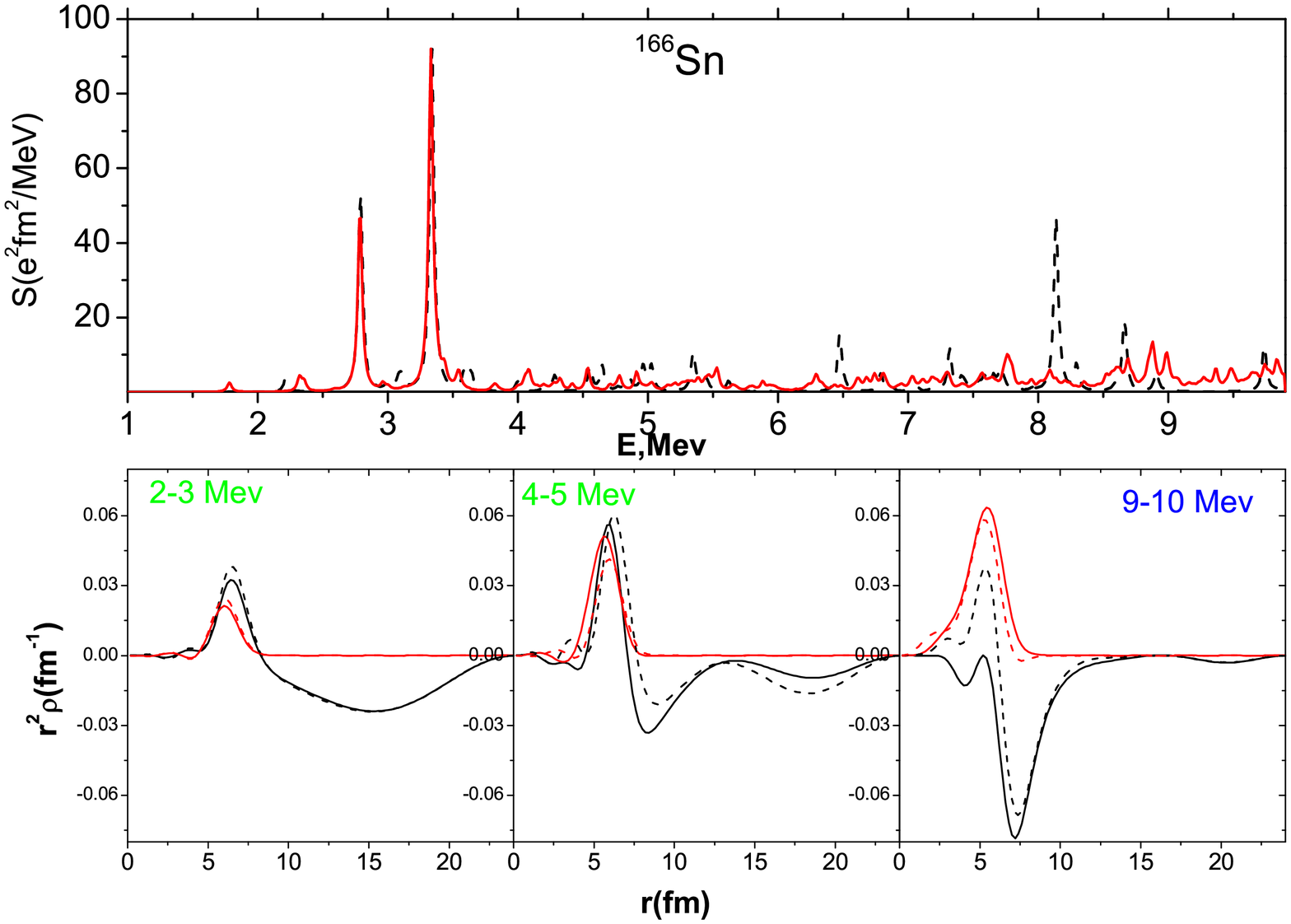}}
 \caption{(Color online) For each of the stable $^{124}$Sn and unstable $^{136}$Sn,  $^{166}$Sn, the upper panel shows the strength function obtained within QRPA (dashed curves) and QTBA (solid curves) in the (0-10)MeV interval  and the lower pannel the corresponding transitional densities for protons~(red curves) and neutrons~(black curves) summed over the  indicated intervals. The smearing parameter is 20~keV. }
\label{fig:trans}
\end{figure}

\begin{figure}
{\includegraphics[width=.95\linewidth,height=0.70\linewidth,angle=-0]{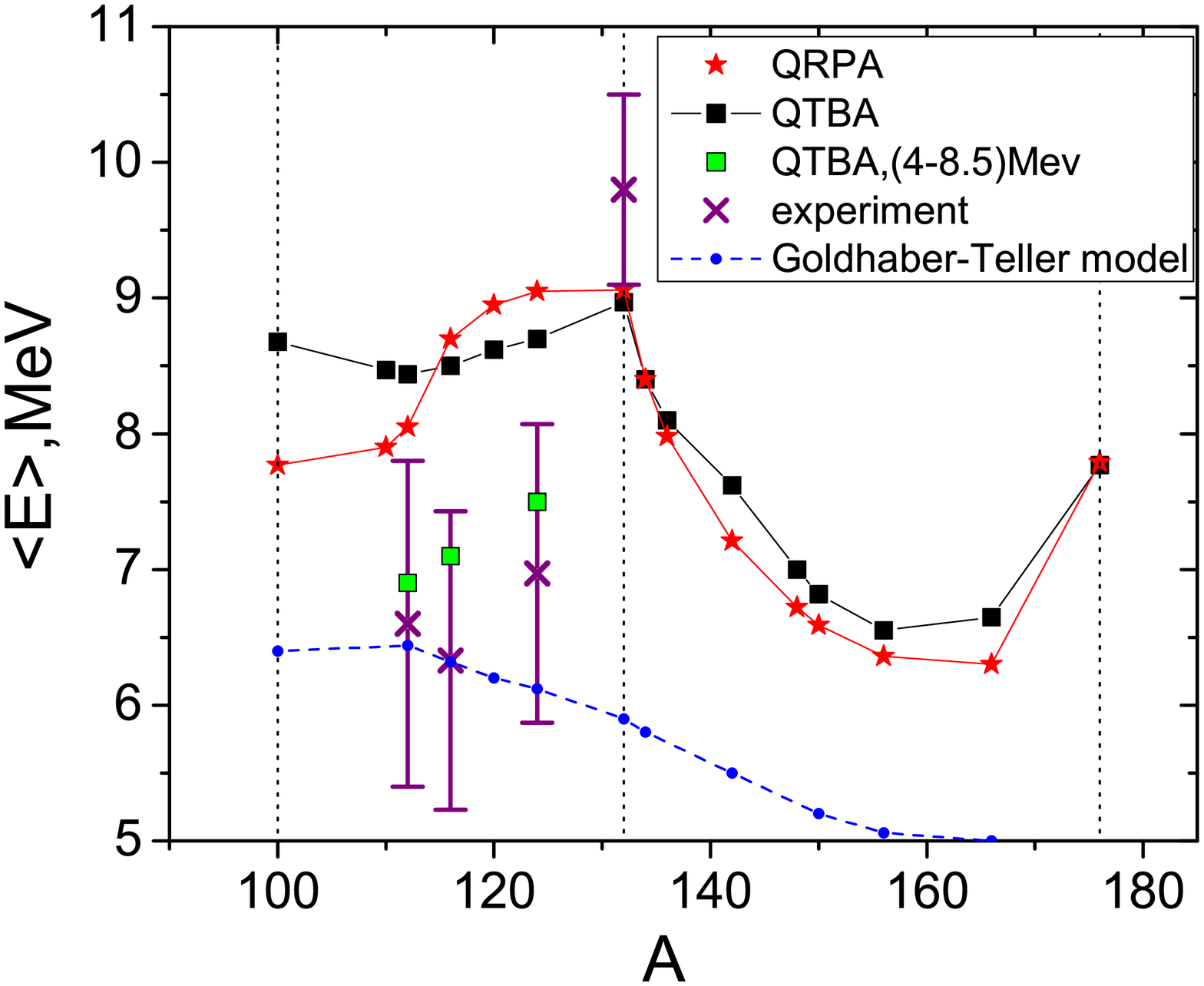}}
\hfill {\includegraphics[width=0.95\linewidth,height=0.70\linewidth,
angle=-0]{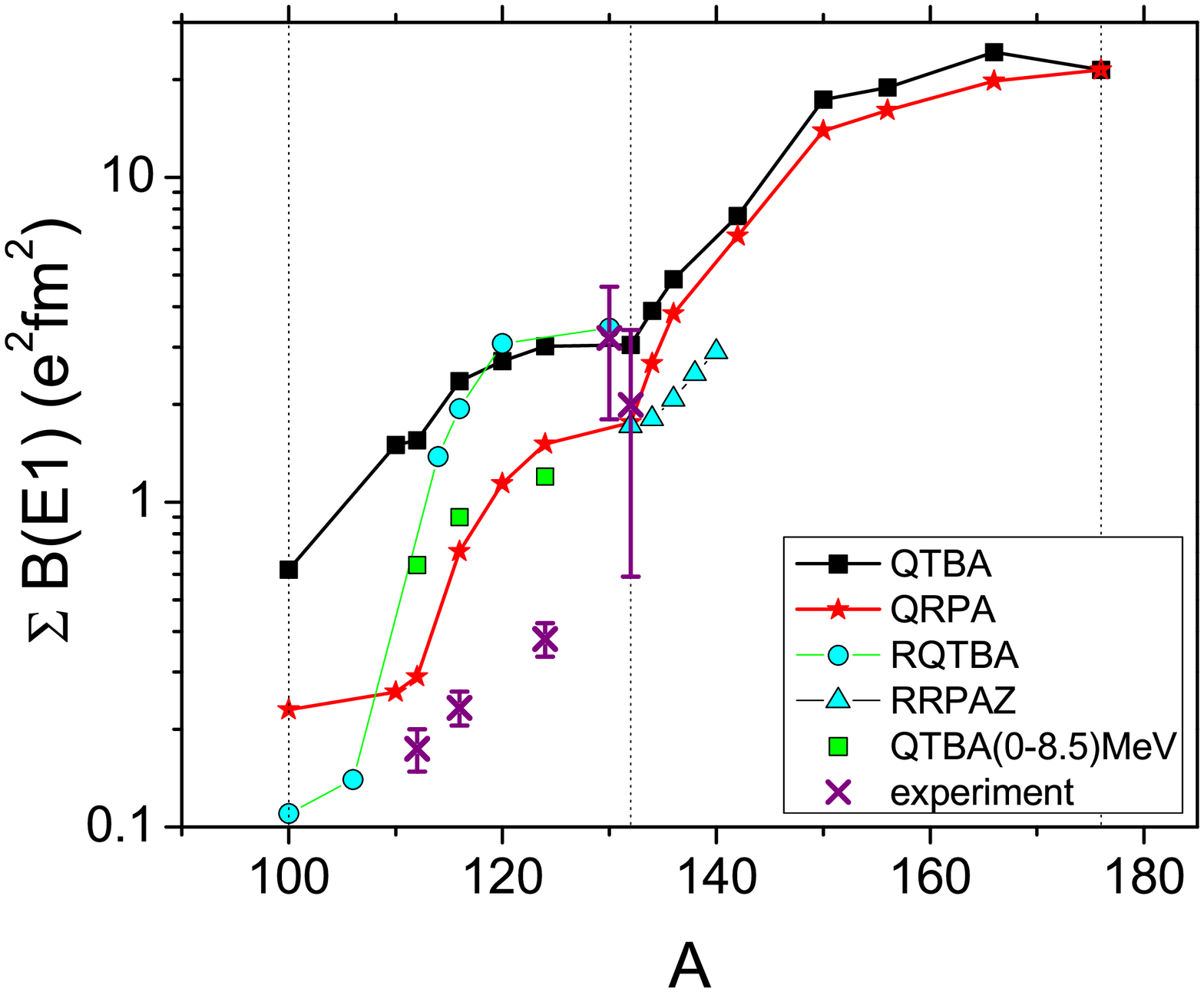}}
 \caption{(Color online) Integral characteristics of $1^{-}$ states
 in the (0-10)MeV interval:
{\it upper panel}- the mean energy versus the atomic mass number $A$;
{\it lower panel}-  $\Sigma B(E1)$ values.
  RQTBA and RQRPAZ  results are taken from Ref.~\cite{ring2009,litvaring1}, respectively.
The results for Goldhaber-Teller model are based on the work~\cite{vanisacker}. Experimental results for $<E>$ and $\Sigma B(E1)$ values are taken from
Refs.~\cite{ponomarev2,zilges07},  respectively.}
  \label{fig:in2}
 \end{figure}

The division of the dipole excitation between the GDR and PDR regions seems rather artificial though the nature of the vibrations is rather different: the low-energy neutron and proton transitional densities are vibrating mainly in phase, while in the GDR energy region these are out of phase~(see for example the case of $^{208}$Pb in Ref.~\cite{tselspeth2007}). But there is no general rule for defining the interval of  pygmy dipole excitations for a given nucleus. We can only visually outline the transition region between the GDR and PDR. We find that for the whole Sn chain this region is rather well described by the  8 to 10 MeV interval, as illustrated in Fig.~\ref{fig:trans}.
This is why we consider here the $(0-10)$~MeV interval for all the nuclei to get a general understanding about the low-lying dipole strength: the resulting mean energies $<E>=E_{1,0}$ and $\Sigma B(E1)$ values are shown in Fig.~\ref{fig:in2}, along with the RQTBA~\cite{litvaring1} and RQRPAZ~\cite{ring2009} results as well as the experimental data  available~\cite{ponomarev2,zilges07}.
We compare our results with these two relativistic approaches as they are the only available calculations of integral characteristics for both the GDR and PDR
in many isotopes of the Sn chain and they have been obtained within a self-consistent scheme. It has to be noted that experimental data is available for $^{112}$Sn, $^{116}$Sn and  $^{124}$Sn isotopes up to the neutron separation energy (the low-lying strength is mostly concentrated in the $(4-8.5)$~MeV interval) while the "pygmy region" for $^{130}$Sn and $^{132}$Sn isotopes is not indicated in Ref.~\cite{zilges07}.

We obtain a reasonable agreement with experiment for the $<E>$ values summed over the $(4-8.5)$~MeV interval (Fig.~\ref{fig:in2}) while the integral strength is a few times larger than the experimental value. A similar behavior was observed in other self-consistent calculations~\cite{ring2009}, see Fig.5. We find some sort of agreement with the experimental $^{132}$Sn  data (for our  $(0-10)$~MeV interval) which gives an integrated strength of the PDR of about (4$\pm$3)\% of the EWSR \cite{newexp}, while our calculation gives 4\% with the PC included and 2\% without. At the same time the calculated and experimental GDR mean energies are rather different and moreover the experimental mean energy is out of the general trend~(Fig.\ref{fig:in1}). Out of our self-consistent calculations for the Sn chain and other  non self-consistent calculations as in Ref.~\cite{tsoneva}, we conclude that the PDR is very model-dependent and probably the force adopted needs some modifications for a simultaneous description of both the GDR and PDR.  It is rather evident also that further experimental investigations are needed as well.

Figs.~\ref{fig:trans}-\ref{fig:in2} demonstrate that there is a distinct difference for  characteristics of a low-lying strength for stable and unstable nuclei. Namely, the PC contribution to the $\Sigma B(E1)$ values, i.e. the difference between QRPA and QTBA predictions, is small for nuclei in the $A>132$ region, while in the $A<132$ region the PC has a rather important impact. Moreover, for nuclei such as $^{112}$Sn, $^{116}$Sn and  $^{124}$Sn (Fig.4, upper panel) the PDR is almost completely defined by complex configurations in the $(4-8.5)$~MeV interval. For the $<E>$ values we also have a "border" at $A=132$: in the $A>132$ region there is almost no PC contribution and a decrease of $<E>$, while in the $A<132$ region this picture is more complicated: the PC contribution  ``corrects'' an  A-dependence of the $<E>$ values which would be expected within the QRPA approach, and in this sense the PC effect is important.

We find that the structure of the PDR excitation spectrum  is very specific to each nucleus. It is hardly a collective mode and can not be described by some systematics like the GDR. For example, the Goldhaber-Teller model adapted for neutron-rich nuclei in Ref.~\cite{vanisacker} gives a rather smooth $A$-dependent behaviour for the PDR mean energy and its strength (Fig.~\ref{fig:in2}) which is not predicted by our calculations and which is not really confirmed by available experimental data. This is easy to understand from the simple Brown-Bolsterly model for the (Q)RPA approach. Indeed, in this model the excitation properties are determined by the structure of single-particle or single-quasi-particle levels. Namely, the larger the difference between the neighbouring levels the more collective the appropriate (Q)RPA 1$^-$-level. From this point of view, the collectivity is determined by the structure of single-(quasi)particle levels and, therefore, the PDR structure is rather specific to each nucleus.
Moreover, recently a  thorough theoretical analysis on the PDR collectivity in $^{132}$Sn was performed in Ref.~\cite{lanza09} which shows that the collectivity is rather weak and only a few particle-hole configurations contribute to the PDR. The authors demonstrated that such contributions are force-dependent and cooperative but not coherent.

\begin{figure}
{\includegraphics[width=.95\linewidth,height=0.70\linewidth,angle=-0]{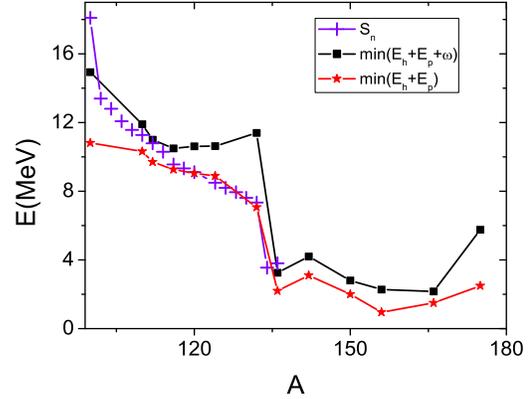}}
\caption{(Color online)
Comparison of the neutron separation energy with the minimal particle-hole energy $(E_{p}+E_{h})$ and the minimal energy ($E_{p}+E_{h}+\omega$, where $\omega$ is the energy of the lowest phonon) for the different Sn isotopes. See text for more details.}
  \label{fig:e1e2w}
\end{figure}

In order to understand better the PC role in the PDR region for stable and unstable nuclei it is useful to consider the
correlation between the neutron  separation energy and the beginning of the low-energy excitation spectrum.
In Fig.~\ref{fig:e1e2w}, we compare the neutron separation energy with the minimal particle-hole energy $(E_{p}+E_{h})$ and the minimal energy $(E_{p}+E_{h}+\omega)$ (where $\omega$ is the energy of the lowest phonon for a given isotope), which approximately determine   the ``beginning'' of the low-lying tail for the (Q)RPA and QTBA models, respectively.
We see clearly that the smaller the neutron separation energy the lower the energy of the first 1$^-$ level.
One can see also a border at $A=132$, i.e a direct correlation and similarity in the beginning of the spectrum within both the (Q)RPA and QTBA for the neutron-rich isotopes.
Due to the small neutron separation energy in the very neutron-rich Sn isotopes a relatively strong low-lying tail of the strength function arises very naturally both within the  QRPA and QTBA~(see Figs.~\ref{fig:st2},\ref{fig:trans}).
So, there is no noticeable difference here between these two approaches. In contrast, for the lighter Sn isotopes~(see Figs.~\ref{fig:st1},\ref{fig:trans})  a considerable contribution of the PC can be observed.

To conclude our analysis of the PDR, we consider transition densities which is now a standard way in investigating the nature of the nuclear excitations. Recently this analysis was performed for some tin isotopes  within the QRPA approach \cite{tsoneva,paarprl,lanza09} and within the relativistic QTBA~\cite{litvaring2}. In Fig.\ref{fig:trans}, we show our self-consistent QRPA and QTBA results for the stable $^{124}$Sn isotope and  the unstable $^{136}$Sn and $^{166}$Sn.
We obtain a rather similar behaviour for $^{136}$Sn and $^{166}$Sn isotopes within the QRPA and  QTBA approaches: below 8 MeV ($^{136}$Sn) and 9 MeV ($^{166}$Sn) the proton and neutron transition densities are in phase in both approaches, i.e. both have an isoscalar character and are clearly dominated by the neutron contribution at the surface. At higher energies they show an isovector behaviour.   Globally, our results are in accordance with  the QRPA results of \cite{tsoneva,paarprl} for stable nuclei. However, we would like to emphasize some major differences between the QRPA and QTBA approaches as far as the transition densities  for the stable $^{124}$Sn isotope are concerned (Fig.\ref{fig:trans}). Physically this corresponds to the fact that the PC gives a considerable contribution to the low-lying strength in stable nuclei, which can clearly be seen in Fig.~\ref{fig:trans} and in Fig.~\ref{fig:st1} for other stable nuclei.

In summary, we find that with the inclusion of the PC the low-lying tail is predominantly of isoscalar nature up to about 8 MeV for all considered Sn isotopes while the ($\approx 8-10$)~MeV interval is a transition region towards the isovector type of excitation which distinguishes the GDR. We also conclude that the inclusion of the PC is necessary to explain the PDR integral properties (including the integrated strength) in stable  isotopes. Moreover it is mostly the PC that contributes below the neutron separation energy. For the $A>132$ nuclei, and especially for unstable neutron-rich nuclei, the PC leads essentially to a redistribution of the PDR strength.

\section{Radiative neutron capture cross sections}
The presence of the PDR in neutron-rich nuclei is of particular interest since, if located well below the neutron separation energy, it can significantly increase the radiative neutron capture cross section and affect the nucleosynthesis of neutron-rich nuclei by the so-called r-process  \cite{gor98,gor02,gor04,arnould07}.  Similarly, the presence of extra strength at low energy in neutron-deficient nuclei can be at the origin of an increase of the radiative proton capture or photoproton emission that take place on the left side of the valley of $\beta$-stability during the so-called rp-process or p-process, respectively \cite{arnould03}.
Since such nucleosynthesis processes involve exotic nuclei that cannot be produced in the laboratory (at least on the neutron-rich side),  only self-consistent calculations can provide a reasonable prediction of their electromagnetic excitation properties. The impact of our newly-derived strength functions on the reaction cross section are discussed below.

\subsection{Comparison between QTBA and QRPA}

\begin{figure}
{\includegraphics[width=.95\linewidth,height=0.7\linewidth,
angle=-0]{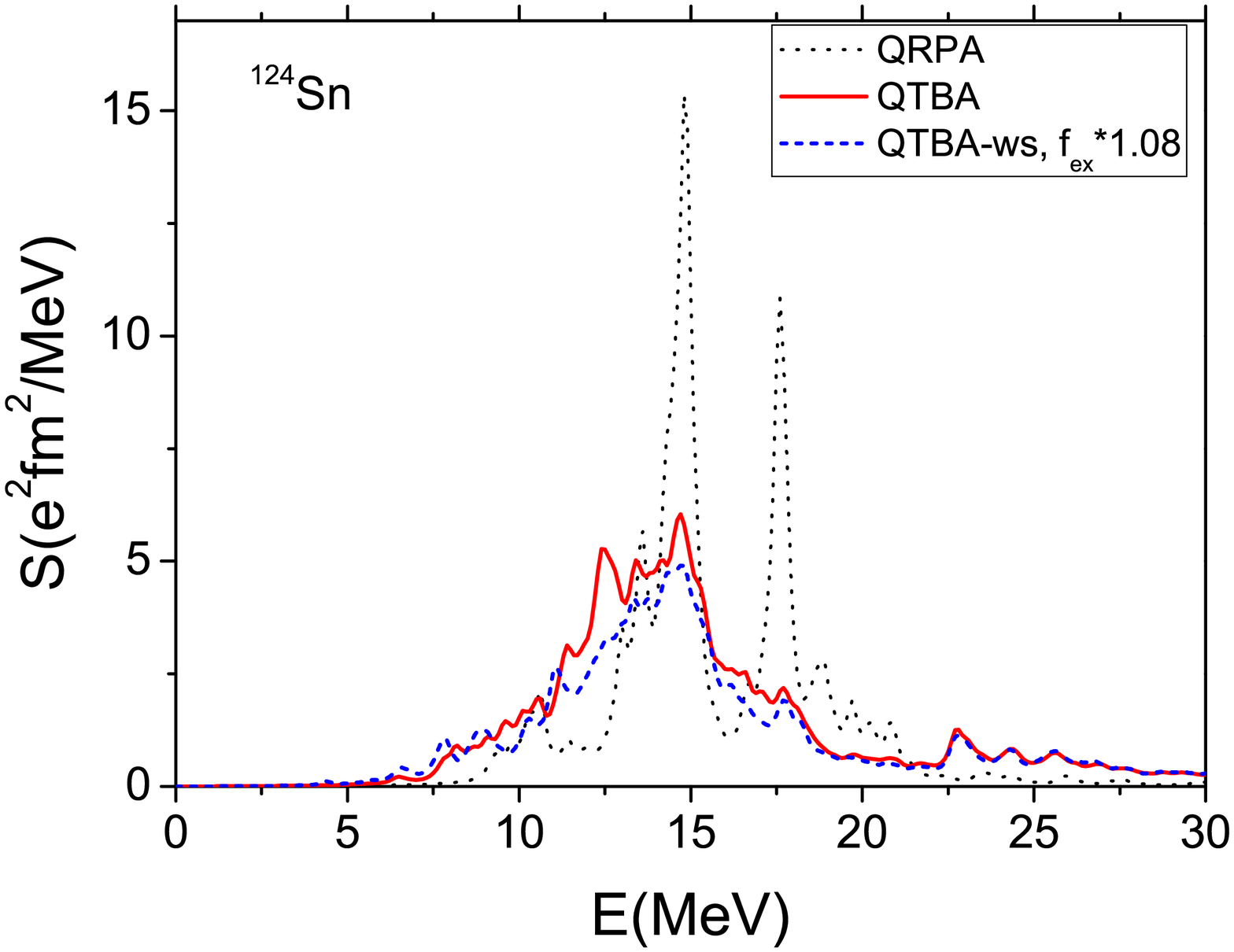}} \hfill
{\includegraphics[width=.95\linewidth,height=0.7\linewidth,
angle=-0]{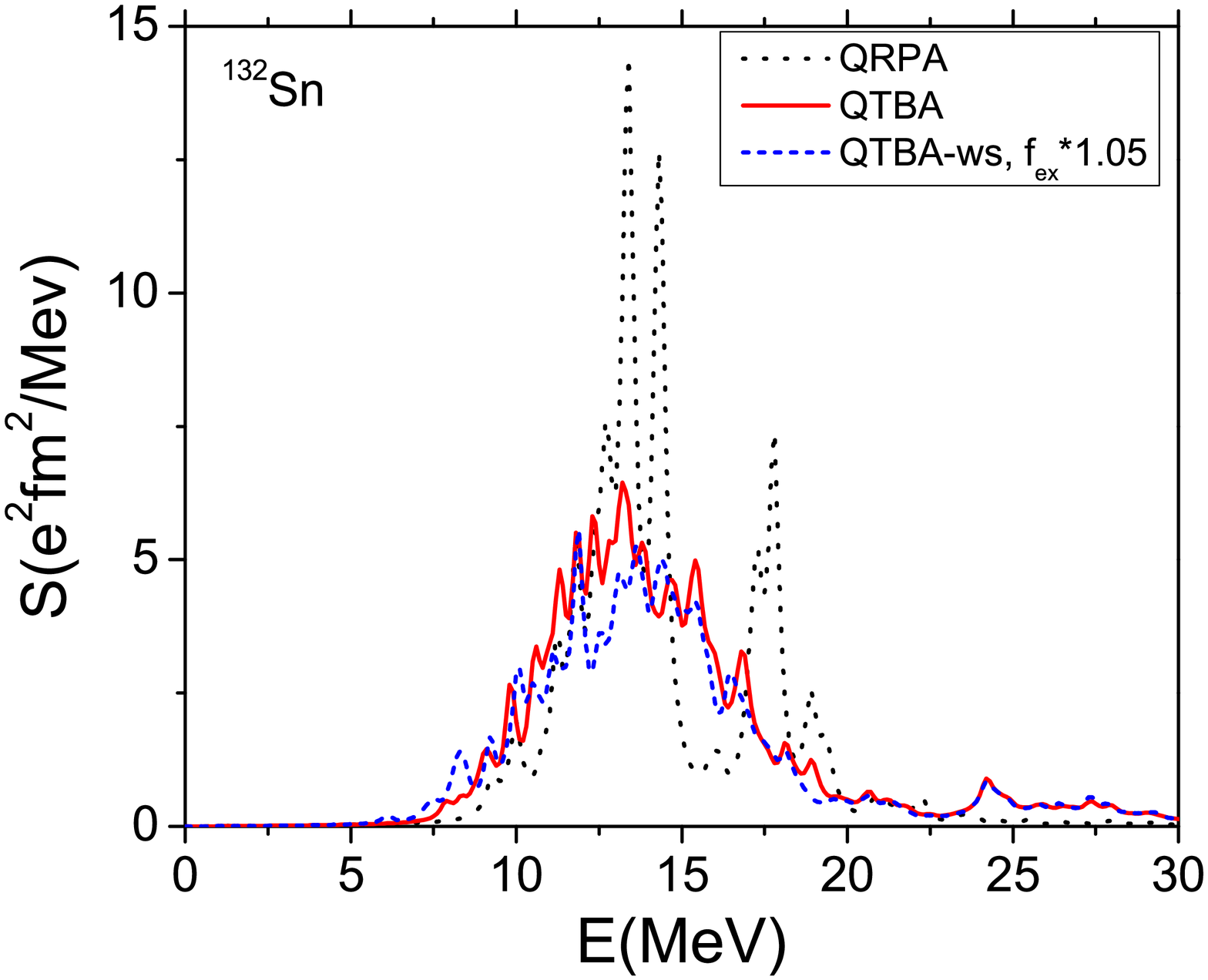}} \hfill
{\includegraphics[width=.95\linewidth,height=0.7\linewidth,angle=-0]{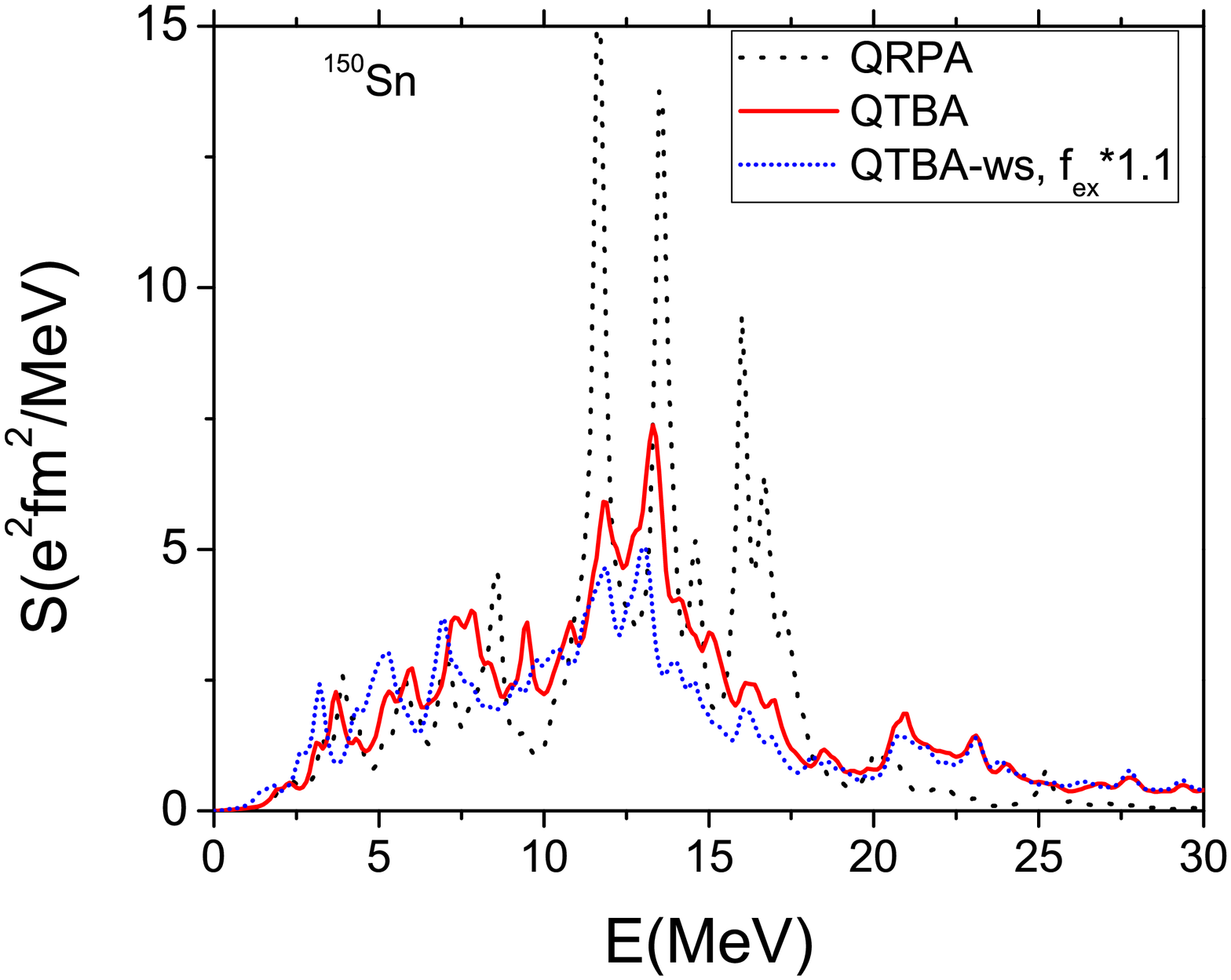}}
 \caption{(Color online)$1^{-}$ strength functions within QRPA~(dotted curves), QTBA~(solid curves) and
  QTBA-ws~(short-dashed curves) for $^{124}$Sn,$^{132}$Sn and $^{150}$Sn isotopes. See text for details}
 \label{fig:rad1}
\end{figure}

\begin{figure}[t!]
{\includegraphics[width=.95\linewidth,height=0.7\linewidth,angle=-0]{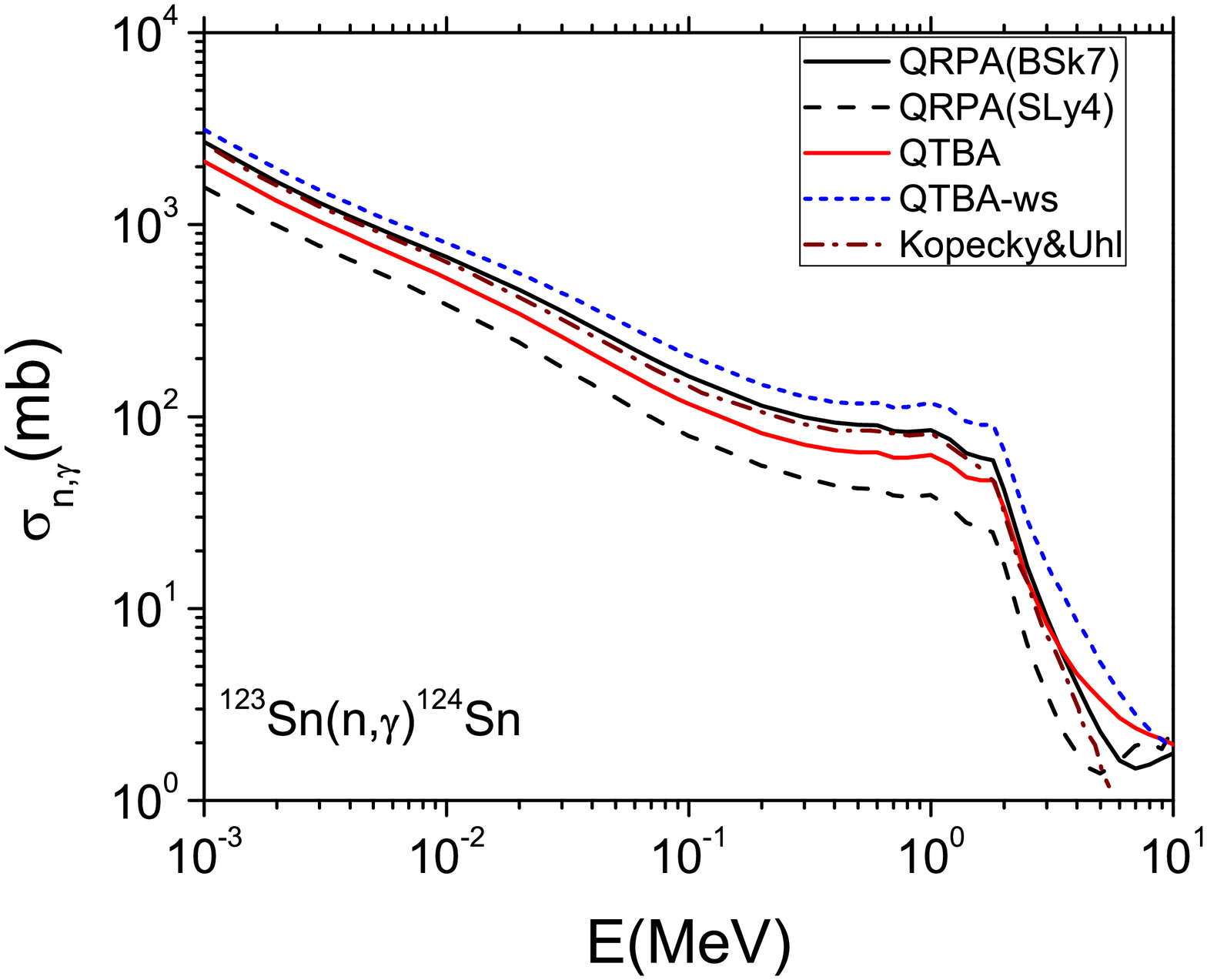}}
{\includegraphics[width=0.95\linewidth,height=0.7\linewidth,
angle=-0]{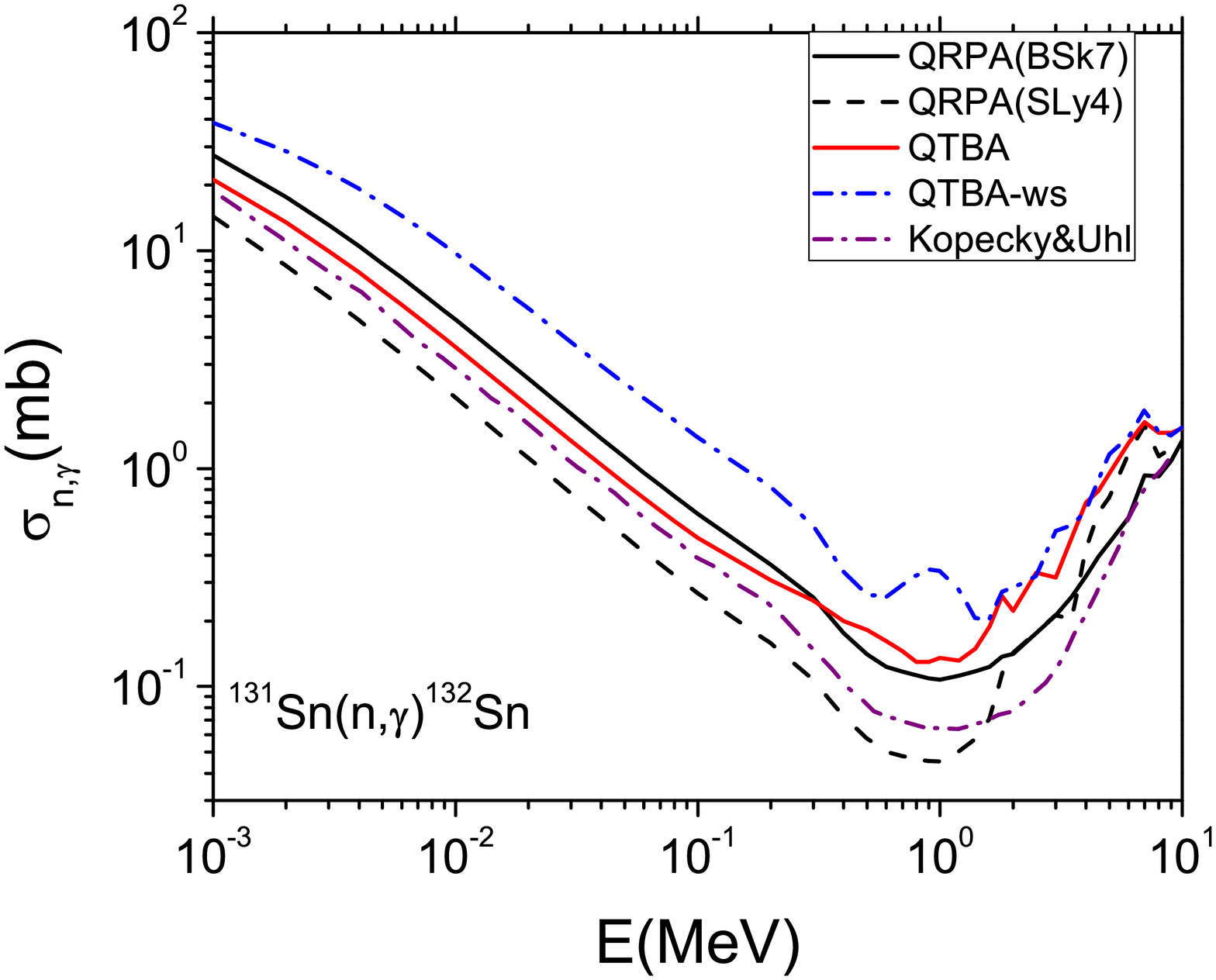}}
\hfill
{\includegraphics[width=.95\linewidth,height=0.7\linewidth,
angle=-0]{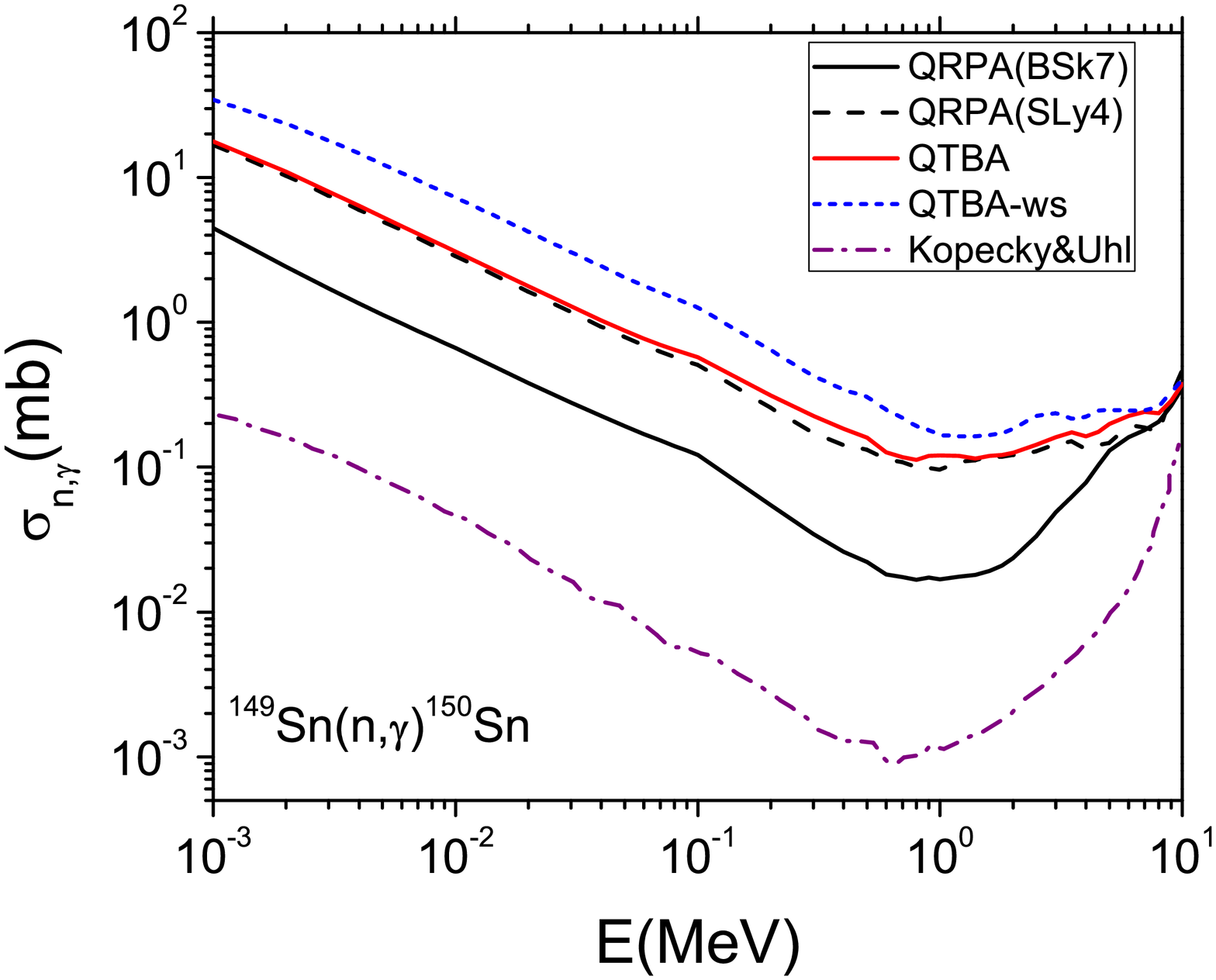}}
 \caption{(Color online) Radiative neutron capture cross sections for  $^{124}$Sn, $^{132}$Sn and $^{150}$Sn
isotopes obtained with the strength functions which were calculated within the QRPA, QTBA and Kopecky-Uhl approaches. See text for details.}
\label{fig:rad2}
\end{figure}

To estimate the impact our new QTBA strength can have on the radiative neutron capture rate of astrophysical interest, the neutron capture cross section is calculated using the reaction code TALYS \cite{go08}. The strength function with and without the PC are included in the calculation of the electromagnetic de-excitation transmission coefficients. The resulting radiative neutron capture cross sections calculated with the strength functions of Fig.~\ref{fig:rad1} are shown in Fig.~\ref{fig:rad2} for the three Sn isotopes.

In Ref.~\cite{avdkoeln2008}, we studied the $^{143}$Nd(n,$\gamma$)$^{144}$Nd cross section using the non-self-consistent ETTFS(QTBA) strength function. Comparison with the QRPA  version showed that the PC inclusion increases the cross section by a factor 2 and improves the agreement with experiment \cite{EXFOR}.
Very recently,  the results of the radiative neutron capture cross sections calculations  within  the self-consistent  relativistic  QTBA were performed for the four tin isotopes \cite{litcross}.

Since the electromagnetic transmission coefficient corresponds to an integral overlap of the de-excitation strength function with the nuclear level density, only the strength function in a restricted energy range below the neutron separation energy play an important role for the estimate of the radiative neutron capture rate \cite{arnould07,rauscher08}. This range corresponds to $\gamma$ energies of typically 2~MeV$< E_{\gamma} <$4~MeV though it may be higher in neutron-deficient nuclei or just before crossing a neutron closed shell.  Therefore if located in this energy range,  the PDR might provide quite a large contribution to the radiative cross section. For neutron-rich nuclei, the mean PDR energy is relatively low and the integrated strength high (Fig.~\ref{fig:in2}), so that the PDR contribution may become significant \cite{gor02,gor04}.

Here we consider  three  different compound Sn nuclei, namely, the stable $^{124}$Sn one and unstable $^{132}$Sn and$^{150}$Sn isotopes. In our earlier calculations \cite{avdkoeln2008,koreya} we  used the microscopically calculated (Q)RPA and QTBA strength functions which have been folded with a Lorentzian in order to reproduce the expected width of the strength function. However, such a procedure tends to smear out the detailed structure of the strength function that may be of interest in the specific energy of relevance (as discussed above). For this reason, we consider here the realistic strength functions without any Lorentzian folding.

We consider two variants of the QTBA calculations  in order to compare the  ETFFS(QTBA) calculations with and without~(QTBA-ws) the subtraction procedure (see the end of Sect.~IIA).
In the variant QTBA-ws,  the isovector part of the calculated effective interaction strength $f_{ex}$ is renormalized  in order to bring (in this approach) the mean energy $E_0$ (Eq.~2) to the QTBA predicted value, i.e to fulfill the condition $E_{0}(QTBA)= E_{0}(QTBA-ws)$. It turns out that the change of the $f_{ex}$ value is  no more than 10\% (more precisely 8, 5 and 10\% for $^{124}$Sn, $^{132}$Sn and  $^{150}$Sn, respectively). We find that  the difference between the QTBA and QTBA-ws strength functions is not large,  both of them differing significantly from the QRPA predictions (see Fig.~\ref{fig:rad1}). The differences between QTBA and QTBA-ws  are essentially found in the redistribution of the strength. This effect is however large enough to impact the radiative neutron capture cross section as shown in Fig.~\ref{fig:rad2}.
In particular, Fig.\ref{fig:rad2} shows that the QTBA-ws gives a larger cross section compared with the QTBA for all the three isotopes although the GDR mean energies are the same for both variants. The comparison with the corresponding RQTBA calculations for $^{131}{\rm Sn}(n, \gamma)^{132}$Sn cross section \cite{litcross} shows, in general, similar behaviour, except for specific energies like those around $E_n=100$~keV for which our cross section is noticeably smaller than the RQTBA one. These deviations may stem from the specific structure of the strength function in the energy range of relevance, as well as from the use of different nuclear ingredients in the cross section calculation, such as nuclear level densities.

In Refs.~\cite{avdkoeln2008,koreya}, where the microscopic strength function is folded with a Lorentzian function, we found that for the stable $^{124}$Sn the cross sections obtained with QRPA and QTBA were almost identical. Comparing the QTBA and QRPA strength functions~(Fig.\ref{fig:rad1}) and cross sections (Fig.\ref{fig:rad2}) allows us to deduce directly the role of the PC (without any interference of an additional Lorentzian smoothing).
In particular, the inclusion of the PC increases the cross sections by a factor 2-3 in the QTBA-ws case (Fig.~\ref{fig:rad2}). In the latter case, the cross section obviously follows the strength function shown in  Fig.~\ref{fig:rad1}, i.e. the extra low-lying strength is responsible for an increase of the reaction cross section (and consequently of the Maxwellian-averaged reaction rate of astrophysical interest) by about a factor of 3 with respect to the  predictions based on the HFB+QRPA calculation with BSk7 Skyrme force~\cite{gor04}. More specifically,  the low-energy $E1$ strength originating from the PC contribution increases the cross section at 0.1~MeV (an energy of relevance for the r-process nucleosynthsesis) by about 30\%  for $^{123}$Sn(n,$\gamma$)$^{124}$Sn for the QTBA and by a factor of about 2 for QTBA-ws. For the  $^{131}$Sn(n,$\gamma$)$^{132}$Sn these figures are even  larger and for $^{149}$Sn(n,$\gamma$)$^{150}$Sn we obtained a similar  effect for the QTBA-ws only.
Thus, in this section we demonstrated the noticeable sensitivity of the radiative neutron capture  cross section with respect to the model (QRPA, QTBA or QTBA-ws) as well as the force  used ~(SLy4, BSk7).

\subsection{Comparison with phenomenological models}
The Lorentzian approach has been widely used for practical applications, though it suffers from shortcomings of various sorts. On the one hand, the location of the GDR maximum energy and width remains to be predicted from some underlying model for each nucleus. For many applications, these properties have often been obtained from a droplet-type model or from experimental systematics  \cite{ripl2}. As shown in Fig.~\ref{fig:in1}, these estimate may differ significantly from the predictions obtained from sounder microscopic models.
In addition, the Lorentzian model tends to overestimate the  $E1$ strength at energies below the neutron separation energy. Different parametrizations or functional forms (including in particular an energy- and temperature-dependent width) have been proposed (see e.g. \cite{ripl2,ku}) to reconcile experimental data in the photon or radiative neutron capture channels, but none of the proposed closed forms can nowadays explain the various trends observed at low energies. Besides the Lorentzian approach cannot provide any predictions on the low-energy PDR, neither on its presence, nor on its characteristics. For this reason, it is of particular interest to analyze to what extent our predictions based on self-consistent microscopic models differ from those used in practical applications.

In Fig.~\ref{fig:rad2}  our results are compared with those obtained with the phenomenological Generalized Lorentzian (GLO) strength function \cite{ku}. For the stable $^{124}$Sn, the GLO cross section is rather similar  to those   obtained within the ETFFS approach, to be exact, this cross section curve is just between the QTBA and QTBA-ws ones, although the strength functions can differ  at low energies below the neutron separation energy (Fig.\ref{fig:rad1}).
However, for  neutron-rich nuclei, such as $^{132}$Sn and $^{150}$Sn, the cross section obtained with the GLO strength on the one hand, and both the QRPA and QTBA, on the other hand, differ, especially  for $^{150}$Sn. As shown in Sect.~II, the main reason lies in the $A$-dependence of  the integral characteristics, but also in the existence of a low-lying strength predicted by the microscopic models. Note that the GLO  parameters used here for $^{150}$Sn correspond to the RIPL2 recommended systematics, i.e. $E_0=14.81$~MeV, $\Gamma =4.47$~MeV and $\sigma_0=341.5$~mb which strongly differ from our microscopic predictions (see Fig.~\ref{fig:in1}).
Fig.~\ref{fig:rad1} also shows the spreading of the strength function down to the lowest energies, i.e in the vicinity of the neutron separation energy, while the GLO model would only provide the tail of the GDR strength at these energies. These comparisons demonstrate the non-applicability of the empirical  systematics and the necessity to make use of self-consistent approaches for  neutron-rich  nuclei. (On the comparisons of M1 resonances with used systematics see Ref.\cite{kako}.)

\section{Conclusion}
The electric dipole strength function has been estimated on the basis of  the ETFFS(QTBA)  model which simultaneously takes into account the (Q)RPA configurations, the more complex 1p1h $\otimes$ phonon or 2 quasiparticle $\otimes$ phonon configurations and the single-particle continuum.
For the long chain of  tin isotopes, the strength functions have been determined within our DTBA approach which is a discretized self-consistent version of the ETFFS(QTBA). The QTBA strengths have been compared with the
(Q)RPA ones which allowed us to study the contribution of the phonon coupling along the whole isotopic chain.

Our conclusions concerning the GDR and PDR properties clearly differ depending on the nuclear region considered, namely  the $A<132$  and $A>132$ regions. More precisely,

\begin{enumerate}
\item for neutron-rich $A>132$ Sn isotopes,  we find, both within QRPA and QTBA,  a significant difference in the A-dependence  of  the GDR  mean energy with respect to the standard phenomenological systematics.
Our Eq.~\ref{eq:eph} gives a new phenomenological systematics;

\item although for  all considered isotopes, the PC contribution to the GDR width is very important quantitatively ,  its contribution to the nuclei of $A>132$ region is relatively  smaller than it is for the $A<132$ nuclei;

\item the PC contribution to the PDR integral characteristics $<E>$ and $\Sigma B(E1)$
 summed over the (0-10) MeV interval is small for the neutron-rich  isotopes;

\item the transition densities in most of the low-energy region are mainly of isoscalar nature  both within the QRPA and the QTBA approaches.  The PC contribution to the transition densities also affects the transition densities, especially in  stable  nuclei.
 Globally, the isoscalar behavior is revealed on the energy interval considered here (1~MeV).
It is not found that the transition density of all individual peaks in the low-energy region is of isoscalar nature. We note also that for these reasons
the QRPA cannot explain quantitatively the isoscalar-.
-isovector splitting of the PDR in the stable $^{140}$Ce observed in the ($\alpha, \alpha' \gamma$) reaction  \cite{paarprl}, see \cite{avekaev2009} as well.

\end{enumerate}

Such a different manifestation of the PC  for nuclei with  $A<132$ and $A>132$  correlates very well with the neutron separation energy. Namely,  the differences are much smaller for neutron-rich nuclei than they are for $A<132$ nuclei.  Just due to this fact  the low-energy parts of the strength functions  in neutron-rich nuclei   are rather similar within the QRPA and QTBA.

The radiative neutron capture cross sections for $^{124}$Sn,$^{132}$Sn and $^{150}$Sn were calculated with the QTBA and QRPA strength functions and shown to be sensitive to the predicted low-lying strength. Significant deviations from the phenomenological GLO approach \cite{ku} are also obtained for the strength functions, and consequently for the neutron capture and photoabsorption cross sections,  for the very neutron-rich isotope $^{150}$Sn. A direct comparison between the QTBA (including the PC) and GLO cross sections shows that the neutron capture cross section on very neutron-rich nuclei may be increased by 2 order of magnitude with respect to the traditional use of phenomenological models. Our results confirm the necessity to use self-consistent microscopic models when dealing with exotic nuclei. Therefore,  nuclear data libraries should not recommend such phenomenological models in this case, but rather point towards newly-developed microscopic large-scale calculations.

The PC and single-particle continuum, which have been included, in addition to the QRPA effects, in our calculations, are necessary ingredients to describe the electric GDR and PDR properly. Nevertheless, it is necessary to use some renormalization  procedures (either by adjusting some interaction parameters or applying the subtraction method) to obtain the correct value of the spurious state energy. In addition, the approach still needs to  include the self-consistency at the level of complex configurations. Therefore, further  developments are needed (see also  \cite{avekaev2009}, where some unsolved issues in the PDR physics are discussed).

\section{Acknowledgments}
We thank Prof.J. Speth for valuable discussions and Dr.N.A. Lyutorovich for the RPA single-particle continuum results in Fig.~2 for magic nuclei $^{132}$Sn and $^{176}Sn$. The work was partly supported by the DFG  grant No. 436RUS113/994/0-1 and  RFBR grant No.09-02-91352NNIOa.


\begin{thebibliography}{100}
\bibitem{VG} V.G. Soloviev, Theory of Atomic Nuclei: Quasiparticles and Phonons, Institute of Physics, Bristol and Philadelphia, USA, 1992.
\bibitem{colo94} G. Colo, Nguyen Van Giai, P.F. Bortignon, R.A Broglia, Phys. Rev.\textbf{C 50} (1994) 1496.
\bibitem{ETFFS2004} S. Kamerdzhiev, J. Speth, and G. Tertychny, Phys. Rep.  \textbf{393},1 (2004)
\bibitem{tse07} V. Tselyaev, Phys. Rev \textbf{C 75}, 024306 (2007)
\bibitem{tsoneva} N. Tsoneva, H. Lenske, Phys. Rev. \textbf{C 77}, 024321 (2008).
\bibitem{sarchi} D.Sarchi, P.F. Bortignon, G. Colo, Phys. Lett. \textbf{B 601} (2004) 27.
\bibitem{litvaring1}E.Litvinova, P. Ring, V.Tselyaev, Phys. Rev. \textbf{C 78}, 014312 (2008).
\bibitem{avd07} A. Avdeenkov,  F. Gruemmer, S. Kamerdzhiev et al., Phys. Lett.  \textbf{B 653}, 196(2007)
\bibitem{kaev1} S. Kamerdzhiev, Bul. Rus. Acad. Sci. Phys, \textbf{61}, 152 (1997).
\bibitem{kaev2} S. Kamerdzhiev, E. Litvinova, Phys. At. Nucl. \textbf{64}, 627 (2001)
\bibitem{kaevliotta1998} S.Kamerdzhiev, R.Liotta, E. Litvinova, V. Tselyaev, Phys.Rev. \textbf{C 58}, 172 (1998)
\bibitem{littsel2007} E.Litvinova, V. Tselyaev, Phys.Rev \textbf{C 75}, 054318 (2007)
\bibitem{rev} N. Paar, D.Vretenar, E. Khan, and G. Colo, Rep. Prog. Phys. \textbf{70}, 691 (2007)
\bibitem{sagawa} H. Sagawa, H. Esbensen, Nucl. Phys. \textbf{A 693}, 448 (2001)
\bibitem{gor02} S.~Goriely, E.~Khan, Nucl. Phys. \textbf{A706}, 217 (2002).
\bibitem{gor04} S.~Goriely, E.~Khan, M. Samyn, Nucl. Phys. \textbf{A739}, 331 (2004).
\bibitem{colo2009} L. Capelli, G. Colo and J. Li, Phys. Rev\textbf{ C 79},054329 (2009).
\bibitem{kneisl} U. Kneisl, N. Pietralla, A. Zilges, J. Phys. G,\textbf{ 32} R217 (2006)
\bibitem{volz} S. Volz, N. Tsoneva, M. Babilon et al., Nucl.Phys. \textbf{A 779}, 1 (2006)
\bibitem{ripl2} T. Belgya, et al., \emph{Handbook for calculations of nuclear reaction data, RIPL-2}, IAEA-Tecdoc-1506 (2006).
\bibitem{avekaev2009} A.V. Avdeenkov and S.P. Kamerdzhiev Phys.At. Nucl. \textbf{72}, 1332 (2009)
\bibitem{gor98} S. Goriely , Phys. Lett.  \textbf{B 436}, 10  (1998).
\bibitem{capote09} R. Capote  et al., Nucl. Data Sheets \textbf{110}, 3107 (2009)
\bibitem{suzuki1990} Y.Suzuki, K.Ikeda, H. Sato, Progr.Theor.Phys. 83, 180 (1990)
\bibitem{wieland}O. Wieland, A. Bracco, F. Camera et.al., Phys. Rev. Lett. \textbf{102}, 092502 (2009).
\bibitem{avdkoeln2008} A. Avdeenkov, S.Goriely S.Kamerdzhiev and G.Tertychny , Gamma-Ray Spectroscopy and Related Topics, 13-th Int. Symp., Cologne Germany 25-29 August 2008, AIP Conf. Proc.\textbf{1090}, p.149.
\bibitem{tert07}G. Tertychny, V. Tselyaev, S. Kamerdzhiev et al.,  Phys. Lett.  \textbf{B 647}, 104 (2007).
\bibitem{colo2001} G.Colo and P.F. Bortignon, Nucl. Phys\textbf{.A696}, 427 (2001).
\bibitem{kaevlit2004} S.P.Kamerdzhiev, E.V. Litvinova, Phys. At. Nucl. \textbf{67}, 183 (2004).
\bibitem{ponomarev2} K. Govaert, F. Bauwens, J. Bryssick et al., Phys. Rev. \textbf{C57}, 2229 (1998).
\bibitem{ponomarev1}R.-D. Herzberg, P. von Brentano, J. Eberth et al.,Phys. Lett. \textbf{B 390}, 49 (1997)
\bibitem{avddubna2009} A.V. Avdeenkov, S. Goriely, S.P. Kamerdzhiev, Phys. At. Nucl. \textbf{73}, 1119 (2010).
\bibitem{lyutor08} N.Lyutorovich, J. Speth, A. Avdeenkov et al., EPJA \textbf{37}, 381 (2008).
\bibitem{bender} M. Bender, G. F. Bertsch, and P.-H. Heenen, Phys. Rev. \textbf{C 73}, 034322
      (2006).
\bibitem{dobach}  M. Kortelainen,  J. Dobaczewski, K. Mizuyama, and J. Toivanen, Phys. Rev.\textbf{ C 77},  064307 (2008).
\bibitem{SLy4} E. Chabanat, P. Bonche, P. Haensel, J.Meyer, R. Schaeffer, Nucl.Phys.\textbf{A 635} (1998) 231.
\bibitem{terasaki1} J. Terasaki, J. Engel,M. Bender et al., Phys. Rev \textbf{C71} (2005) 034310
\bibitem{bennaceur} K.Bennaceur and J. Dobaczewski,Comp.Phys.Comm,\textbf{168} (2005) 96
\bibitem{speth} S.O.Backman,A.D.Jackson, J.Speth, Phys.Lett, \textbf{B56} (1975),209
\bibitem{krewald1977} S.Krewald, V.Klempt, J.Speth, A. Faessler, Nucl. Phys. \textbf{A281}, 166 (1977).
\bibitem{shlomo} S. Shlomo and G. Bertsch, Nucl. Phys. \textbf{A243}, 507 (1975).
\bibitem{kiae} E.E. Saperstein, A.V. Tolokonnikov and S.A. Fayans  Preprint \textbf{IAE 2571} (in Russian), Moscow 1975
\bibitem{kst93} S. Kamerdzhiev, J. Speth, G. Tertychny and V.Tselyaev, Nucl.Phys. \textbf{A555} (1993) 90.
\bibitem{tselyaev00} V.I. Tselyaev, Bull. Rus. Acad. Sci. Phys. \textbf{64} (2000) 434.
\bibitem{yadfiz1983} S.P. Kamerdzhiev, Sov. J. Nucl.Phys.\textbf{ 38}, 188 (1983).
\bibitem{aumann05}A. Adrich, A.Klimkiewicz, M.Fallot et al.,Phys. Rev. Lett. {\bf 95}, 132501 (2005)
\bibitem{BSk5} M.Samyn, S. Goriely, P.-H. Heenen, J.M. Pearson, F. Tondeur, Nucl. Phys. \textbf{A 700}, 142 (2002).
\bibitem{SkM*} J. Bartel, P.Quentin, M. Brack, C. Guet, H.-B. Hakansson, Nucl. Phys. \textbf{A 386}, 79 (1982).
\bibitem{tselspeth2007} V.Tselyaev, J.Speth, F.Gruemmer, S. Krewald, A. Avdeenkov, E. Litvinova and G. Tertychny, Phys.Rev \textbf{C75}, 014315 (2007).

\bibitem{ring2009} D.Pena Arteaga, E.Khan and P.Ring, Phys. Rev. \textbf{C79}, 034311 (2009).
\bibitem{ringsch} P. Ring, P. Schuck, The nuclear many-body problem, Springer, 1980.
\bibitem{zilges07} B. Ozel, J. Enders, P. von Neumann-Cosel et al., Nucl. Phys. \textbf{A 788}, 385 (2007).
\bibitem{newexp} A. Klimkiewicz, P. Adrich, K. Boretzky et al., Nucl. Phus. \textbf{A 788}, 145c (2007).
\bibitem{vanisacker} P. Van Isacker, M.A. Nagarajan and D.D. Warner. Phys. Rev.\textbf{ C45}, R13 (1992).
\bibitem{lanza09} E.G. Lanza, F.Catara, D. Gambacurta et al., Phys. Rev. \textbf{C79} 054615 (2009).
\bibitem{litvaring2}E.Litvinova, P. Ring, V. Tselyaev, K. Langanke, Phys. Rev. \textbf{C79}, 054312 (2009).
\bibitem{paarprl} N.Paar, Y.F. Niu, D. Vretenar, and J. Meng, Phys. Rev. Lett. \textbf{103}, 032502 (2009)
\bibitem{arnould07} M. Arnould, S. Goriely, T. Takahashi, Phys. Rep. \textbf{450}, 97 (2007).
\bibitem{arnould03} M. Arnould, S. Goriely, Phys. Rep. \textbf{384} (2003)  1
\bibitem{go08} S.Goriely, S.Hilaire, A. J. Konig, Astron. Astroph. \textbf{487}, 767 (2008)
\bibitem{EXFOR} EXFOR library: http://www-nds.iaea.or.at/exfor
\bibitem{litcross} E.Litvinova, H.P. Loens, K.Langanke, G. Martinez-Pinedo, T.Rauscher, P. Ring, F.-K. Thielemann, V. Tselyaev, Nucl. Phys. \textbf{A 823}, 26 (2009).
\bibitem{rauscher08} T.Rauscher,   Phys. Rev \textbf{C78}, 032801(R) (2008).
\bibitem{koreya} A. Avdeenkov, S. Goriely, S. Kamerdzhiev, Proc. Int. Conf. on Nuclear Data (26-30.04.2010, Koreya) , to be published;arXiv:1103.2630v1[nucl-th].
\bibitem{ku} J.Kopecky, M. Uhl, Phys. Rev.\textbf{ C41}, 1941 (1990).
\bibitem{kako} S. P. Kamerdzhiev and S. F. Kovalev, Phys. At. Nucl. \textbf{69}, 418 (2006).
\end{thebibliography}
\end{document}